\documentclass[review]{elsarticle}

\usepackage{amsmath}
\usepackage{amsfonts}
\usepackage{amssymb}
\usepackage{graphics}
\usepackage{graphicx}

\usepackage{multirow}
\usepackage{bm}

\usepackage{color} 
\definecolor{mygreen}{RGB}{28,172,0} 
\definecolor{mylilas}{RGB}{170,55,241}
\usepackage{subcaption} 
\newcommand{\sgn}{{\rm sgn}}
\newcommand{\sech}{{\rm sech}}

\journal{...}

\begin{document}

\begin{frontmatter}

\title{Nonlinear simulation of wave group attenuation due to scattering in broken floe fields}

\author[a1]{Boyang Xu}
\ead{boyangxu@udel.edu}

\author[a1]{Philippe Guyenne\corref{cor1}}
\ead{guyenne@udel.edu}

\address[a1]
{Department of Mathematical Sciences, University of Delaware, DE 19716, USA}

\cortext[cor1]{Corresponding author}


\begin{abstract}
Direct phase-resolved simulations are performed to investigate the propagation and scattering of nonlinear ocean waves in fragmented sea ice.
The numerical model solves the full time-dependent equations for nonlinear potential flow coupled with 
a nonlinear thin-plate representation of the ice cover, and neglects dissipative processes.
The two-dimensional setting with incident wave groups on deep water is considered,
in view of applications to wave attenuation along transects of the marginal ice zone.
A spatially-varying weight is assigned to the surface pressure so that irregular distributions of ice floes 
can be directly specified in the physical domain.
For various wave regimes and floe configurations, a local wave spectrum across the ice field is computed
and then least-squares fitted to extract a spatial attenuation rate as a function of wave frequency.
A general increase with frequency is found, consistent with typical predictions from linear theory.
However, a non-monotonic behavior around a certain frequency is also observed, especially in cases with a highly fragmented ice cover.
Comparison to field data shows that this result closely resembles the roll-over phenomenon
as reported from Arctic Ocean experiments.
The key role played by nonlinear interactions in the roll-over development is highlighted.
\end{abstract}


\end{frontmatter}


\section{Introduction}

The rapid decline of summer ice extent that has occurred in the Arctic Ocean over recent decades is a striking illustration of the major transformations
experienced by the polar regions due to global warming.
While there is no doubt that warmer temperatures have been a crucial factor in these transformations,
it is now recognized that the action of ocean waves and their increased activity play an aggravating role in controlling sea-ice morphology.
Ice melting and proliferation of open water promote further wave growth over increasing fetches,
thus allowing long waves to propagate larger distances into the ice field.
Waves can break up the sea ice and cause it to become more fragmented,
which then increases their capacity to further penetrate and damage the ice cover.
In turn, the presence of sea ice affects the wave dynamics, with various effects depending on the ice type.
These processes are especially prevalent in the marginal ice zone (MIZ) which is the fragmented part of the ice cover closest to the open ocean.
Unfortunately, such information has not been factored into previous climate predictions and this is now becoming a pressing issue.
It is only recently that operational wave forecasting models have begun to be tested with refined parameterizations for wave-ice interactions 
\cite{cheng17,db13,wbsdb13b}.

A problem of particular interest to oceanographers is the representation of wave attenuation by sea ice,
in view of applications to the MIZ.
Two different mechanisms usually contribute to this effect:
(i) scattering by ice floes or other inhomogeneities of the ice cover,
which is a conservative process that redistributes energy in all directions,
and (ii) dissipative processes which may be attributed to various causes,
e.g. turbulence and friction beneath the ice layer, inelastic collisions and breakup of ice floes.
Uncertainties about the actual mechanisms for wave dissipation in sea ice,
and about the relative importance between conservative and non-conservative processes,
have prompted a surge of research activity on this subject in recent years.
Currently, the general consensus is that dissipation is dominant in frazil and pancake ice fields \cite{nm99,rogers16},
while scattering is the main contributor to wave attenuation in broken floe fields
when the floe size is on the order of the wavelength \cite{km08,wsgcm88}.

In this regard, the development of mathematical models has mostly relied on linear theory assuming potential flow,
and can be divided into two main categories:
(i) discrete-floe models where individual floes with possibly distinct features are explicitly resolved in a simple geometry \cite{bs12,km08,msb16},
and (ii) continuum models where the heterogeneous ice field is viewed as a uniform material
with effective rheological properties including viscosity or viscoelasticity \cite{cgg19,dd02,ws10}.
Type (i) is especially designed to investigate scattering and typically requires solving
a boundary value problem involving multiple regions associated with the floe field.
Type (ii) enables the derivation of an exact algebraic expression for the dispersion relation governing plane waves in the effective medium. 
Wave attenuation is encoded in the complex roots of this dispersion relation and is controlled by constant parameters.
Because there are no intrinsic values for these effective parameters, increasing attention has been paid to type (ii) models 
over the last few years to test and calibrate them against experimental data \cite{cheng17,desanti18,mms15}.
Recent reviews on theoretical models of type (i) and (ii) can be found in \cite{s19,s20}.

Ocean waves however are inherently nonlinear and, from the perspective of global warming,
it is also expected that powerful storms with associated large waves will become more frequent and have greater impact on sea ice.
Reports of intense waves-in-ice events have highlighted limitations of linear theory \cite{m03}.
For example, Kohout et al. \cite{kwdm14} measured wave heights of over $3$ m in the Antarctic Southern Ocean
and observed that such large waves attenuate at a much slower rate than predicted by the commonly assumed exponential decay.
As a result, these waves can propagate and maintain ice-breaking potential hundreds of kilometers into the MIZ.

Despite significant progress over the past decade, the nonlinear theory is still in its infancy.
Efforts have mainly focused on the analysis and simulation of flexural-gravity waves in homogeneous sea ice,
using nonlinear potential-flow theory for the underlying fluid combined with linear or nonlinear
thin-plate theory for the floating ice \cite{bmf09,dkp19,hi98,mvbw11,pd02}.
A compelling thin-plate formulation in this situation is the one recently proposed by Plotnikov and Toland \cite{pt11}
based on the special Cosserat theory of hyperelastic shells,
which yields a nonlinear and conservative expression for the bending force.
This allows the full governing equations to be written in canonical Hamiltonian form,
thus extending the classical approach of Zakharov \cite{z68} for water waves to the hydroelastic problem.
Plotnikov and Toland's idea has been adopted in subsequent work by other investigators,
and has produced a number of results ranging from direct numerical simulation 
to weakly nonlinear modeling in various asymptotic regimes \cite{gp12,gp14,mw13}.
The main goal of these nonlinear studies has been to characterize localized traveling waves,
either freely evolving or driven by a moving load on the ice sheet, in the absence of dissipative effects.

Numerical or theoretical developments concerning the nonlinear problem of wave propagation in fragmented sea ice
are even more scarce. This is a particularly challenging problem which remains largely unexplored.
A few numerical models based on the Navier--Stokes equations have been developed 
to simulate fluid flows in the presence of isolated rigid floating bodies \cite{osvbck16}, 
including wave overwash of a floating plate \cite{nbst20}.
For larger-scale applications, Lagrangian sea-ice models like those based on the discrete element method
can reproduce ice breakage and the subsequent formation of cracks and leads \cite{h16}, or the dynamics of colliding floes \cite{hcs19}.
In this context, the wave excitation is specified either via simple parameterization or via coupling with an independent waves-in-ice module.
Phase-averaged forecasting models for the wave spectrum may also include components
to describe wave attenuation by sea ice (due to scattering and dissipation).
As mentioned above, they have been tested with more options for such processes in recent years.
While their parameterizations for wave-ice interaction are typically derived from linear theory,
nonlinear energy transfer is allowed via the usual mechanism of $4$-wave resonant interactions.
Phase-resolved simulations of nonlinear wave propagation in fragmented sea ice
have been conducted by Guyenne and P\u ar\u au \cite{gp17} using a potential-flow solver combined with
a mixed continuum-piecewise extension of Plotnikov and Toland's thin-plate formulation.
Focusing on the shallow-water regime and ignoring dissipative effects, 
these authors provided an examination of solitary wave scattering and attenuation in various floe configurations.
Despite peculiarities related to solitary waves, some general analogy was found between their numerical results 
and previous observations from experiment or theory.

Pursuing this line of inquiry, the present paper is an extension of Guyenne and P\u ar\u au's work in several important ways.
Our new contributions include:
\begin{itemize}
\item[(1)] Consideration of incident wave groups in the two-dimensional deep-water regime.
Several sets of measurements from the Arctic MIZ have revealed that groupiness is a common feature of the wave field
in open-water and ice-covered conditions \cite{crmb15,gmt21,tgrca19}.
While the scattering process is now well understood from a theoretical viewpoint in the linear setting,
with predictions that compare reasonably well to experimental data, it is not clear to what extent
these results would be applicable in the nonlinear setting where many different types of waveforms exist.
\item[(2)] Inclusion of inertial effects in our thin-plate model as these may be relevant when the wavelength is much longer than the floe size.
Their computation is quite nontrivial in this nonlinear time-dependent framework
and must be accommodated as part of the numerical scheme to solve the evolution equations.
\item[(3)] Estimation of the spatial attenuation rate as a function of wave frequency
from direct phase-resolved simulations in various wave regimes and floe configurations.
This calculation is intensive because it involves averaging results over multiple random realizations
of each floe configuration for each grid point in the ice field.
It also requires evaluating an integral over a long time interval for each of these grid points to produce a local wave spectrum
and then least-squares fitting to extract the decay rate.
\item[(4)] Comparison with measurements of attenuation rate from the Arctic and Antarctic MIZ.
We find cases where our numerical estimates agree qualitatively well with the field data
over the range of frequencies measured, and they even provide some extrapolation
at higher frequencies where data are not available.
We also test these numerical results against predictions from existing linear theories.
Our computations confirm the dominant contribution from scattering to the process of wave attenuation in broken floe fields.
\item[(5)] Account of a non-monotonic behavior of decay rate that resembles the roll-over effect
as observed in Arctic MIZ measurements.
Continuum viscoelastic models or discrete scattering models have generally been unable to predict this phenomenon.
Possible explanations that have been proposed include wind forcing, nonlinear interactions or instrument noise \cite{lkdwgs17,ph96,thmkk21}.
While we are unable to pinpoint the exact cause, our numerical results highlight the key role played by nonlinearity 
in the roll-over development, without the need for any external forcing.
To our knowledge, this is the first time that such a roll-over has been revealed from direct phase-resolved simulations
of nonlinear ocean waves propagating in fragmented sea ice.
\end{itemize}

The remainder of this paper is organized as follows.
Section 2 presents the mathematical formulation of this two-dimensional hydroelastic problem,
including reduction of the governing equations to a lower-dimensional system via the Dirichlet--Neumann operator.
It also recalls the main features of the model for fragmented sea ice.
Section 3 outlines the numerical methods for space discretization and time integration,
and describes the incident wave conditions.
Numerical results on wave attenuation for various wave regimes and floe configurations are then shown,
including a description of wave profiles and the estimation of attenuation rates.
Finally, a comparison with field observations is provided
together with a discussion on the roll-over phenomenon in cases where it occurs.

\section{Mathematical formulation}

\subsection{Governing equations}

We consider the motion of an inhomogeneous ice sheet lying over a two-dimensional fluid of infinite depth.
Dissipative processes are neglected in this study.
The fluid is assumed to be incompressible and inviscid, and the flow to be irrotational.
For this introductory presentation of the mathematical formulation, we first consider a continuous homogeneous ice sheet.
It is modeled as a thin elastic plate according to the special Cosserat theory of hyperelastic shells in Cartesian coordinates $(x,y)$, 
with the $x$-axis positioned along the bottom of the ice sheet at rest and the $y$-axis directed vertically upward \citep{pt11}. 
The vertical displacement of the ice relative to $y = 0$ is denoted by $y = \eta(x,t)$. 
Based on the Euler equations for potential flow, the fluid velocity potential $\Phi(x,y,t)$ satisfies the Laplace equation
\begin{equation}
\nabla^2 \Phi=0 \,, \qquad \mbox{for} \qquad x \in \mathbb{R} \,, \quad -\infty < y < \eta(x,t) \,.
\label{laplace}
\end{equation}
The nonlinear boundary conditions at $y = \eta(x,t)$ are the kinematic condition
\begin{equation}
\partial_t \eta + (\partial_x \Phi) (\partial_x \eta) = \partial_y \Phi \,,
\label{kine}
\end{equation}
and the dynamic (or Bernoulli's) condition
\begin{equation}
\partial_t \Phi + \frac{1}{2} \left|\nabla \Phi \right|^2 + g \eta + \frac{P_i}{\rho} = 0 \,,
\label{dyna}
\end{equation}
with $P_i = P_1 + P_2$ where
\begin{equation} \label{pressure}
P_1 = \rho_i h \partial_t^2 \eta \,, \qquad
P_2 = \sigma \left( \partial_s^2 \kappa + \frac{1}{2} \kappa^3 \right) \,.
\end{equation}
The parameters involved are the acceleration due to gravity $g = 9.81$ m s$^{-2}$, 
the fluid density $\rho = 1025$ kg m$^{-3}$, the ice density $\rho_i = 917$ kg m$^{-3}$, 
the ice thickness $h$ and the ice flexural rigidity $\sigma$.
The latter is taken to be
\[
\sigma = \frac{E h^3}{12 (1 - \nu^2)} \,,
\]
where $E = 6$ GPa and $\nu = 0.3$ denote Young's modulus and Poisson's ratio for the ice respectively.
A characteristic value for ice thickness in the MIZ is $h = 1$ m \cite{mms15}.
The thin-plate approximation is usually valid when the characteristic wavelength is much larger than the characteristic ice thickness.
The pressure $P_i$ exerted by the ice sheet onto the fluid surface has two components.
The acceleration term $P_1$ in \eqref{pressure} is associated with inertial motion,
while the nonlinear bending force due to plate elasticity is given by
\begin{equation} \label{cosserat}
\partial_s^2 \kappa + \frac{1}{2} \kappa^3 = \frac{1}{\sqrt{1 + (\partial_x \eta)^2}} \partial_x 
\left[ \frac{1}{\sqrt{1 + (\partial_x \eta)^2}} \partial_x \left( \frac{\partial_x^2 \eta}{\big[ 1 + (\partial_x \eta)^2 \big]^{3/2}} \right) \right] 
+ \frac{1}{2} \left( \frac{\partial_x^2 \eta}{\big[ 1 + (\partial_x \eta)^2 \big]^{3/2}} \right)^3 \,,
\end{equation}
where 
\[
\kappa = \frac{\partial_x^2 \eta}{\big[ 1 + (\partial_x \eta)^2 \big]^{3/2}} \,,
\]
represents the mean curvature at any point of the fluid-ice interface and $s$ is the arclength along this interface.
In anticipation of the numerical methods described later, 
periodic boundary conditions will be imposed in the horizontal direction.
The system of equations is closed with a vanishing-flux boundary condition
\begin{equation}
\partial_y \Phi \to 0 \,, \qquad \mbox{as} \qquad y \to -\infty \,.
\label{bottom}
\end{equation}
We also assume that the elastic plate is not pre-stressed and neglect plate stretching.

An initial motivation for using \eqref{cosserat} to model the bending force is that it is both nonlinear and conservative,
meaning that it can be associated with some potential energy that is conserved over time.
This has some appeal because it can fit within the classical Hamiltonian formulation of the water wave problem \cite{z68},
which has implications for analytical and numerical investigation.
Previous work based on such a formulation for nonlinear flexural-gravity waves 
propagating along a uniform ice sheet can be found in \cite{gp12,gp14,mw13}.
However, because we will ultimately consider an inhomogeneous setting, 
the conservative property is not of relevance in this study.

\subsection{Dirichlet--Neumann operator} 

For computational purposes, the Laplace problem \eqref{laplace}--\eqref{bottom} can be reduced to
a lower-dimensional system by introducing $\xi(x,t) = \Phi(x,\eta(x,t),t)$, the trace of the velocity potential 
on the boundary $y = \eta(x,t)$, together with the Dirichlet--Neumann operator (DNO)
\[
G(\eta) : \xi \longmapsto (-\partial_x \eta,1)^\top \cdot \nabla \Phi \big|_{y=\eta} \,,
\]
which is the singular integral operator that takes Dirichlet data $\xi$ on $y = \eta(x,t)$, 
solves the Laplace equation \eqref{laplace} for $\Phi$ subject to \eqref{bottom}, 
and returns the corresponding Neumann data (i.e. the normal fluid velocity there).

With this at hand, the full nonlinear equations of motion \eqref{kine}--\eqref{dyna} can be rewritten as
\begin{eqnarray} \label{eq-eta}
\partial_t \eta & = & G(\eta) \xi \,, \\
\partial_t \xi & = & -g \eta - \frac{P_i}{\rho} - \frac{1}{2} (\partial_x \xi)^2 
+ \frac{\big[ G(\eta) \xi + (\partial_x \eta) (\partial_x \xi) \big]^2}{2 \big( 1 + (\partial_x \eta)^2 \big)} \,,
\label{eq-xi}
\end{eqnarray}
which is a closed dynamical system in terms of the boundary variables $\eta$ and $\xi$ alone.
As shown in \cite{cs93}, the DNO can be expressed via a convergent Taylor series expansion
\begin{equation} \label{series}
G(\eta) = \sum_{j=0}^{\infty} G_j(\eta) \,,
\end{equation}
where each term $G_j$ is homogeneous of degree $j$ in $\eta$ and can be determined recursively.
For example, the first three contributions to \eqref{series} are given by
\begin{eqnarray*}
G_0 & = & |D| \,, \\
G_1 & = & D \eta D - G_0 \eta G_0 \,, \\
G_2 & = & \frac{1}{2} \big( G_0 D \eta^2 D - |D|^2 \eta^2 G_0 \big) - G_0 \eta G_1(\eta) \,,
\end{eqnarray*}
where the operators $G_0$ and $D = -i \, \partial_x$ may be viewed as Fourier multipliers.
In this framework, the function $\eta(x,t)$ representing the upper moving boundary is assumed to be a single-valued graph of $x$
and thus overturning waves (with a multi-valued profile) cannot be described.

The dispersion relation for linear traveling wave solutions of \eqref{eq-eta}--\eqref{eq-xi} 
about the fluid at rest ($\eta,\xi = 0$) is 
\begin{equation} \label{lindisp}
\frac{\omega^2}{g - \frac{\rho_i h}{\rho} \omega^2 + \frac{\sigma}{\rho} k^4} = k \,,
\end{equation}
where $\omega$ and $k$ denote the angular frequency and wavenumber respectively (assuming $k, \omega > 0$).
In the absence of ice ($\sigma = 0$ and $\rho_i = 0$ or $h = 0$),
Eq. \eqref{lindisp} reduces to the linear dispersion relation 
\[
\omega^2 = g k \,,
\]
for deep-water gravity waves, whose group speed is
\[
c_g^{(o)} = \partial_k \omega = \frac{1}{2} \sqrt{\frac{g}{k}} \,.
\]
In the absence of inertial effects (say, for a long uniform ice sheet), Eq. \eqref{lindisp} simplifies to
\[
\omega^2 = g k + \frac{\sigma}{\rho} k^5 \,,
\]
and the corresponding group speed behaves like
\begin{eqnarray} \label{speed}
c_g^{(i)} = \frac{g + 5 \frac{\sigma}{\rho} k^4}{2 \omega} & \simeq & c_g^{(o)} \left( 1 + \frac{9 \sigma}{2 g \rho} k^4 + \dots \right) \,,
\quad \mbox{as} \quad k \to 0 \,, \\
& \simeq & 5 c_g^{(o)} \sqrt{\frac{\sigma}{g \rho}} k^2 \,, \quad \mbox{as} \quad k \to +\infty \,, \nonumber
\end{eqnarray}
which suggests that wave groups travel faster in pack ice than in open water ($c_g^{(i)} > c_g^{(o)}$ for any $k$).

In the following, Eqs. \eqref{eq-eta}--\eqref{eq-xi} are non-dimensionalized using the quantities
$\ell = (\sigma/\rho g)^{1/4}$ and $\tau = (\sigma/\rho g^5)^{1/8}$ as characteristic length and time respectively \cite{bmf09}, 
so that $g = 1$, $\sigma/\rho = 1$ and $\rho_i h/\rho = 0.06$ in dimensionless units.
Note that the typical dimensional values of parameters introduced earlier yield $\ell = 15.3$ m and $\tau = 1.2$ s, 
which lie in the spectrum of wind-driven gravity waves.
For convenience, we will retain the same notations but, unless stated otherwise, 
it is now understood that the values of any variable or parameter are dimensionless.

\subsection{Model of fragmented sea ice}

Having in mind the problem of wave scattering in the MIZ, we borrow the same idea as proposed in \cite{gp17} 
to mimic a fragmented ice sheet consisting of an irregular array of ice floes.
To make the present paper sufficiently self-contained, we recall the main features of this strategy.
The array of ice floes is modeled as a spatial distribution of ``icy'' and ``wet'' areas where the surface pressure $P_i$
is switched ``on'' and ``off'' respectively.
This amounts to assigning a spatial weight $f(x)$ to $P_i$, whose amplitude varies between $0$ (open water) and $1$ (pack ice).
To comply with the underlying continuum formulation of the governing equations and with the global interpolation property
of the numerical methods that we will employ to solve these equations (see next section), 
the transition between the two phases is prescribed smooth but steep enough to clearly identify the individual floes.
A tanh-like profile is typically used to describe such a transition at the edges of each floe.

In this two-dimensional situation, the total horizontal length $L_c$ of the ice field and the total number $N_i$
of constitutive floes are given parameters.
The algorithm first generates a regular array of $N_i$ identical floes of individual length $L_i$
which are uniformly spread over $L_c$.
To produce a more irregular arrangement, each floe is then shifted by a distance $\gamma L_i/2$ relative to its initial center,
where $\gamma$ is a random number uniformly distributed between $-1$ and $1$.
If the leftmost or rightmost floe happens to be shifted outside of the ice field, its exterior part would be trimmed off.
Similarly, if two neighboring floes happen to overlap after this shift, they would be merged into a longer floe.
These effects are not viewed as detrimental because they contribute to making the ice cover more inhomogeneous.

\begin{figure}[t!]
\centering
\includegraphics[width=1.1\linewidth]{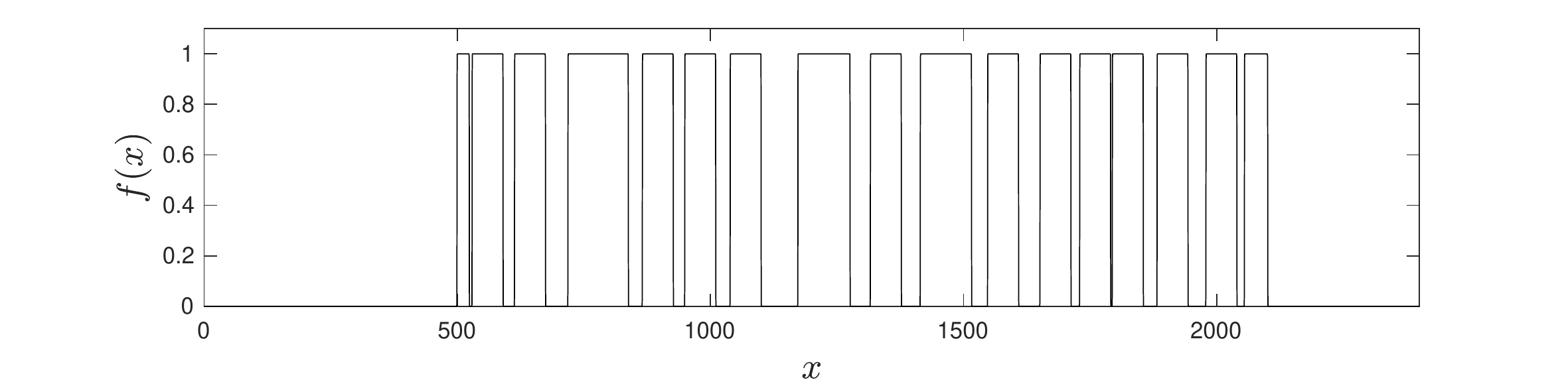}
\caption{One realization of the spatial weight $f(x)$ assigned to the surface pressure $P_i$.
The ice cover spans a distance $L_c = 1600$ between $x = 500$ and $x = 2100$,
and consists of $N_i = 17$ floes whose average length is $L_i = 60$.}
\label{floe_Ni20_Li60_r1}
\end{figure}

The floe distribution is assumed here to be fixed in time, which implies that there is no feedback from waves to floes
and such frictional effects as floe-floe collisions are neglected.
Furthermore, the floes are specified in such a way that they coincide with the water surface and bend in unison with it,
hence free-edge boundary conditions are not considered.
Figure \ref{floe_Ni20_Li60_r1} portrays one realization of the spatial profile $f(x)$ over the entire domain 
$0 \le x \le L$ ($L = 2400$). In this example, there are $N_i = 17$ floes (originally $N_i = 20$) of average length $L_i = 60$ 
which are randomly distributed over $500 \le x \le 2100$ (hence $L_c = 1600$). 
The ice floes are represented by the level sets $f(x) = 1$, and it can be seen that some of them have been merged together.

We recognize that this is a rather heuristic procedure for simulating fragmented sea ice,
as it is meant to be an idealization of the actual problem. A major advantage is that it easily fits within 
the nonlinear time-dependent reformulation \eqref{eq-eta}--\eqref{eq-xi} of the governing equations for fluid motion,
and thus lends itself well to efficient and accurate computations as discussed below.
Preliminary tests of this methodology when applied to incident solitary waves on shallow water show encouraging results in \cite{gp17}.
An objective of the present study is to further assess its applicability by treating the case of incident wave groups on deep water,
which is of relevance to wave-ice interactions in the ocean \cite{crmb15,gmt21,tgrca19}.

\section{Numerical results}

\subsection{Numerical methods and incident wave conditions}

Equations \eqref{eq-eta}--\eqref{eq-xi} are solved numerically following a high-order spectral approach 
which is outlined here. More details on its implementation and validation can be found in \cite{cs93,gp12,gp17}.
Discretization in space is accomplished via a pseudo-spectral method based on the fast Fourier transform.
The computational domain spans the interval $0 \le x \le L$ with periodic boundary conditions 
and is divided into a regular grid of $N$ collocation points.
Applications of spatial derivatives or Fourier multipliers are performed in the Fourier space,
while nonlinear products are calculated in the physical space.
For example, if we wish to apply the zeroth-order operator $G_0 = |D|$ to a function $\xi$ in the physical space,
we first transform $\xi$ to the Fourier space, multiply the diagonal operator $|k|$ with the Fourier coefficients of $\xi$,
and then transform back to the physical space.

The DNO is approximated by a truncation of its Taylor series \eqref{series} for which a small number $M$ of terms 
(typically $M < 10 \ll N$) is sufficient to achieve highly accurate results, 
by virtue of its analyticity properties ensuring spectral convergence.
The number $M = 4$ is selected based on previous extensive tests \cite{gn07,xg09},
and is also deemed sufficient here because we expect wave attenuation as a result of scattering by sea ice, 
in which case the solution becomes less nonlinear and thus does not require a higher value of $M$ to be accurately resolved.
Time integration is performed in the Fourier space so that the autonomous linear terms $G_0 \xi$ and $-g \eta$
can be solved exactly by the integrating factor technique, thus reducing any possible stiffness of the system.
The nonlinear terms are integrated in time using a fourth-order Runge--Kutta scheme with constant step size $\Delta t$.

A novelty in the present model, as compared to \cite{gp12,gp14,gp17}, is the inclusion and numerical treatment of the acceleration term $P_1$.
Inertial effects may be appreciable when the wavelength is much longer than the floe size.
Note that $P_1$ depends nonlinearly on $\eta$ and $\xi$ as it is related to the kinematic boundary condition \eqref{eq-eta},
and is also not autonomous because it involves the weighting function $f(x)$.
Accordingly, this term is treated as a nonlinear term in the implementation of the fourth-order Runge--Kutta scheme.
Its computation is quite nontrivial in the time stepping process since $P_1$ contains a second-order time derivative of $\eta$
and appears in the evolution equation for $\xi$.
Consequently, the method of order reduction is not suitable and $P_1$ must be evaluated directly as part of \eqref{eq-xi}.
After substituting $\partial_t^2 \eta = \partial_t \big( G(\eta) \xi \big)$, we approximate it by first-order finite differences
\[
\partial_t^2 \eta(t_m) \simeq \frac{\big( G(\eta) \xi \big)(t_m) - \big( G(\eta) \xi \big)(t_n)}{t_m - t_n} \,,
\]
for any $t_m$ such that $t_n < t_m \le t_{n+1}$ (where $t_n$ and $t_{n+1} = t_n + \Delta t$ are two regular consecutive time steps).
The $x$-dependence is omitted here for convenience.
This approximation of $\partial_t^2 \eta$ was deemed to be a good compromise between accuracy and simplicity,
fitting well into an explicit time integrator of \eqref{eq-eta}--\eqref{eq-xi}.
It is considered reasonable for our purposes in view of the small dimensionless parameter ($= 0.06$)
associated with $P_1$, and provided that $\Delta t$ is chosen sufficiently small.

For this nonlinear problem of wave scattering by sea ice, care is taken to prescribe a sufficiently accurate traveling wave solution 
to the water wave system (Eqs. \ref{eq-eta}--\ref{eq-xi} with $P_i = 0$) as incident wave conditions.
Otherwise, a too crude approximation or any arbitrary condition would tend to quickly disperse, 
which may then lead to overestimation of the decay rate.
Furthermore, it is preferable to specify an incident wave that is spatially localized to accommodate
the periodic boundary conditions imposed by the pseudo-spectral method,
and to minimize possible issues related to undesirable wave reflection or transmission.
In this deep-water setting, we use a weakly nonlinear wave group (or wave packet) as given by the Stokes expansions
\begin{eqnarray*}
\eta(x,0) & = & \frac{1}{2\omega_0} \partial_x \Phi + A \cos(\theta) 
+ \frac{1}{2} k_0 A^2 \cos(2 \theta) + \frac{1}{2} A (\partial_x A) \sin(2 \theta) + \frac{3}{8} k_0^2 A^3 \cos(3 \theta) 
+ \dots \nonumber \\
\xi(x,0) & = & \Phi + \Big[ \frac{\omega_0}{k_0} A \sin(\theta)
+ \frac{\omega_0}{2 k_0^2} (\partial_x A) \cos(\theta) - \frac{3 \omega_0}{8 k_0^3} (\partial_x^2 A) \sin(\theta)
- \frac{1}{8} \omega_0 k_0 A^3 \sin(\theta) \Big] e^{k_0 \eta} + \dots 
\end{eqnarray*}
with contributions from up to the third harmonics \cite{cgs21}, where
\[
\Phi = i \frac{\omega_0}{2} \sgn(D) A^2 \,, \qquad \partial_x \Phi = -\frac{\omega_0}{2} |D| A^2 \,,
\] 
and
\[
A(x) = a_0 \, \sech \big( \sqrt{2} \, \varepsilon \, \theta \big) \,, \qquad \theta = k_0 (x - x_0) \,.
\]
The incident wave parameters are the amplitude $a_0$, carrier wavenumber $k_0$, angular frequency 
$\omega_0 = \sqrt{g k_0}$ and position $x_0$ of the central (highest) crest.
The quantity $\varepsilon = k_0 a_0$ is a measure of the incident wave steepness.
The function $A(x)$ provides a representation of the wave group envelope, and the phase is controlled by $\theta$.
Fixing $k_0$, the larger $\varepsilon$, the taller and narrower the wave group.
We checked numerically beforehand that, in the absence of ice, such a solution propagates 
without change of shape and speed of its envelope for the parameter values under consideration.

Although the incident wave is localized, the subsequent scattering by ice floes will inevitably produce small-amplitude waves
that radiate backward and forward from the main pulse.
In view of long-time simulations, to prevent these radiative waves from contaminating the advancing solution 
by re-entering the domain due to the periodic boundary conditions,
we also specify absorbing sponge layers through an additional pressure term of the form
\begin{equation} \label{sponge}
-\frac{\beta(x) G(\eta) \xi}{\sqrt{1 + (\partial_x \eta)^2}} \,,
\end{equation}
on the right-hand side of \eqref{eq-xi}, which damps outgoing waves near both ends of the domain and away from the ice cover.
The tunable coefficient $\beta(x)$ is nonzero only over a small region near $x = \{ 0, L \}$ and, similar to $f(x)$,
a tanh-like profile is used to represent its localized spatial behavior.

\subsection{Description of wave profiles}

The main objective of this study is to examine wave attenuation due to scattering by ice floes 
via direct time-dependent simulations in the nonlinear deep-water regime.
To quantify this attenuation, we exploit the weakly nonlinear character of the incident wave conditions
(which is further weakened during the subsequent evolution).
We thus use linear theory as a reference, assuming an exponential behavior with distance traveled through sea ice.
Linear predictions have been tested with reasonable success against field measurements \cite{dd02,nm99,wsgcm88}
and offer a convenient way to estimate the spatial attenuation rate as a function of wave frequency.
Although we choose localized wave groups to be the incident waves, which is advantageous for computational purposes, 
we point out that such solutions possess a well-defined oscillatory structure with wavenumber $k_0$ 
and thus can be assigned a respective frequency (unlike the solitary waves on shallow water that were prescribed in \cite{gp17}).

As part of this investigation, we perform a series of tests and explore the dependence of results 
on various wave and ice parameters such as incident wave steepness $\varepsilon$, 
ice concentration $C \simeq N_i L_i/L_c$ and ice fragmentation $F \simeq N_i$.
The quantities $N_i L_i/L_c$ and $N_i$ should be viewed as average values of $C$ and $F$ (for a regular floe arrangement)
since their actual values may slightly vary from one realization to another as a result of floe trimming or merging
by the randomization procedure.
Because of the high computational cost associated with our direct time-dependent simulations,
only a small ensemble of such realizations will be generated for each set of parameter values,
and only a limited number of wave regimes and floe configurations will be inspected.

For the purposes of our discussion (in particular to allow for an even comparison among the various cases considered),
we fix the proportions of the computational domain to $L = 2400$ and $L_c = 1600$,
and we set the numerical parameters to $N = 16384$ ($\Delta x = 0.14$) and $\Delta t = 0.002$.
The ice field is confined to the interval $500 \le x \le 2100$ and the incident wave group is initially centered at $x_0 = 400$.
These parameter values were selected based on computational and physical considerations after extensive trials, 
as a good compromise for accurate and manageable simulations while enabling the observation of wave attenuation 
over a sufficiently long distance. We point out that these computations are intensive (even more so than in \cite{gp17})
since they require solving the nonlinear nonlocal equations \eqref{eq-eta}--\eqref{eq-xi} 
in a very large domain and over an extended period, using fine resolutions in both space and time.
Because of these numerical constraints, it is understood that our simulations are not meant to resolve
the vast range of length and time scales associated with wave-ice interaction in the MIZ.

For the tests, we focus on two distinct wave steepnesses $\varepsilon = 0.05$ (low steepness, weak nonlinearity)
and $\varepsilon = 0.2$ (higher steepness, stronger nonlinearity),
and four different floe configurations defined by $(N_i,L_i) = (124,4)$, $(124,8)$, $(20,60)$, $(1,1600)$.
These floe arrangements correspond to average ice concentrations $C = 0.31$, $0.62$, $0.75$, $1.00$ respectively.
They also display separate levels of ice fragmentation:
the ice cover for $(N_i,L_i) = (124,4)$ and $(124,8)$ is highly fragmented, consisting of many small floes, 
while that for $(N_i,L_i) = (20,60)$ is less fragmented, with fewer but larger floes.
The special case $(N_i,L_i) = (1,1600)$ represents a single long floe spanning the entire ice field.
For each of these floe configurations, given a value of $\varepsilon$, we run simulations for a range of values of $(a_0,k_0)$
in order to obtain estimates of the spatial attenuation rate as a function of wave frequency, 
similar to data sets reported from laboratory or field experiments.
We understand that the wave steepness may not be constant across the spectrum of wave frequencies
that has been typically measured in such studies.
Here we do so for convenience by assuming that, for each data set, all wavenumbers (hence all frequencies)
correspond to the same wave steepness. 
Otherwise, mixing up various regimes together would make the analysis more complicated.
Furthermore, distinguishing data sets (of decay rate vs. frequency) according to different wave steepnesses
will allow for a meaningful comparison between them.

For $\varepsilon = 0.05$, we considered a set of eleven wavenumbers distributed over the range $0.24 \le k_0 \le 2.20$
(accordingly the range of wave amplitudes $a_0 = \varepsilon/k_0$ is $0.02 \le a_0 \le 0.21$).
For $\varepsilon = 0.2$, we chose the set of eleven wavenumbers to be in the range $0.12 \le k_0 \le 2.00$
(accordingly the range of wave amplitudes is $0.10 \le a_0 \le 1.67$).
The choice of these values for $k_0$ is dictated on one hand by the spatial and temporal discretizations
(to ensure that the shortest wavelength $\lambda_0 = 2\pi/k_0$ in the range and its associated wave period $2\pi/\omega_0$ 
are sufficiently well resolved), and on the other hand by the proportions of the computational domain
(to ensure that the broadest incident wave group allowed by $\varepsilon$ and $k_0$ can fit well into
the open-water area on the left side of the ice field).
Note also that, for each floe configuration $(N_i,L_i)$, with the exception of the single-floe case $(N_i,L_i) = (1,1600)$,
and for each pair of wave parameters $(a_0,k_0)$ given a value of $\varepsilon$,
we perform computations for ten different random realizations of the fragmented ice sheet, as described in Sec. 2.3.
To give the reader an idea about the computational cost entailed in this work,
the run time of a simulation for each set of $(a_0,k_0,N_i,L_i)$ up to $t = 1500$ was about a week 
on the HPC community cluster at the University of Delaware. 
It took more than a year and half to complete all the runs for the various sets of parameter values as mentioned above 
(without counting preliminary tests on the simulation setup).

\begin{figure}[t!]
\centering
\begin{subfigure}[h]{1.1\textwidth}
\includegraphics[width=\linewidth]{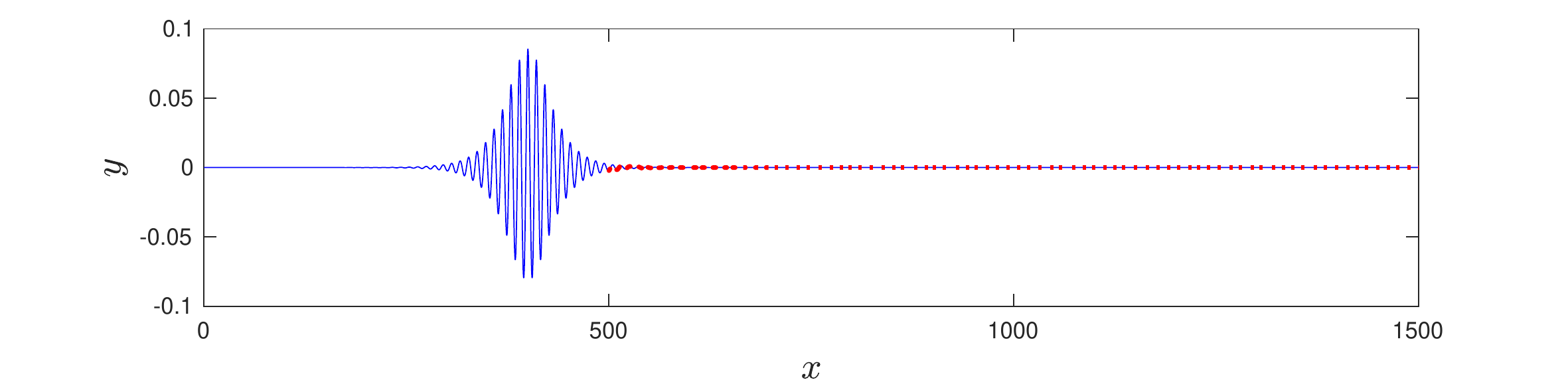} 
\caption{$t = 0$, $N_i = 124$, $L_i = 4$}
\end{subfigure}
\begin{subfigure}[h]{1.1\textwidth}
\includegraphics[width=\linewidth]{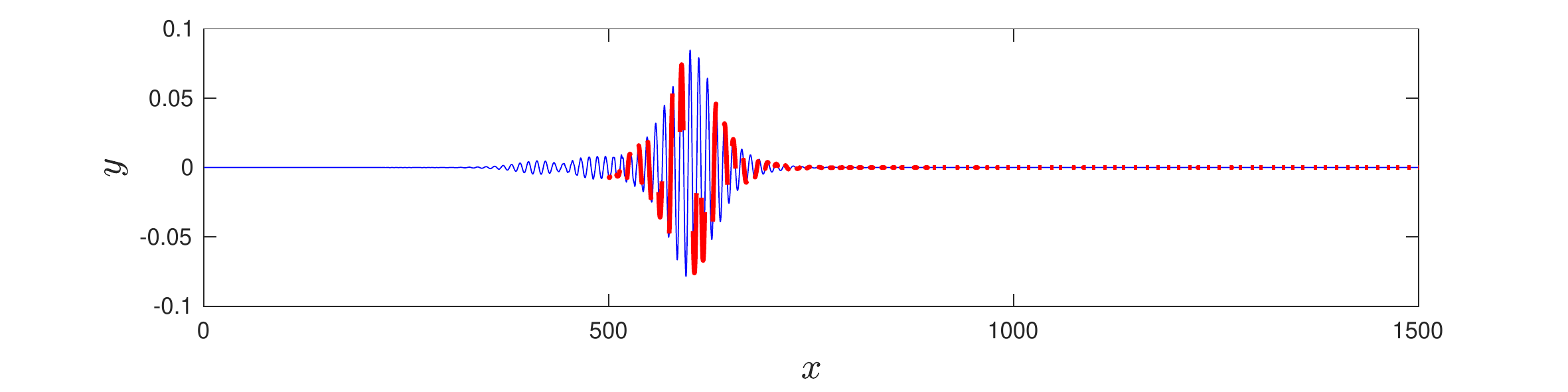}
\caption{$t = 300$, $N_i = 124$, $L_i = 4$}
\end{subfigure}
\begin{subfigure}[h]{1.1\textwidth}
\includegraphics[width=\linewidth]{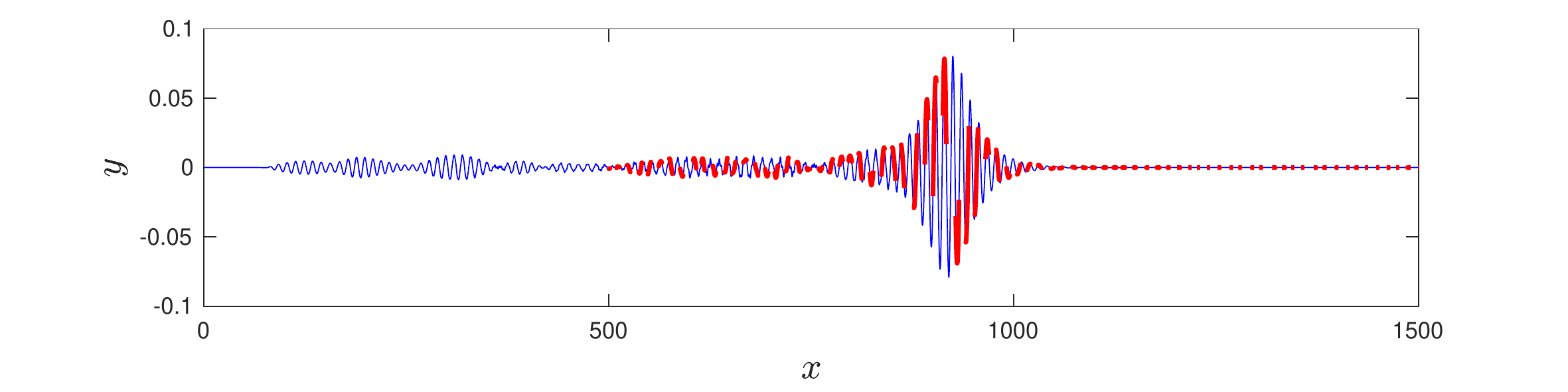}
\caption{$t = 750$, $N_i = 124$, $L_i = 4$}
\end{subfigure}
\begin{subfigure}[h]{1.1\textwidth}
\includegraphics[width=\linewidth]{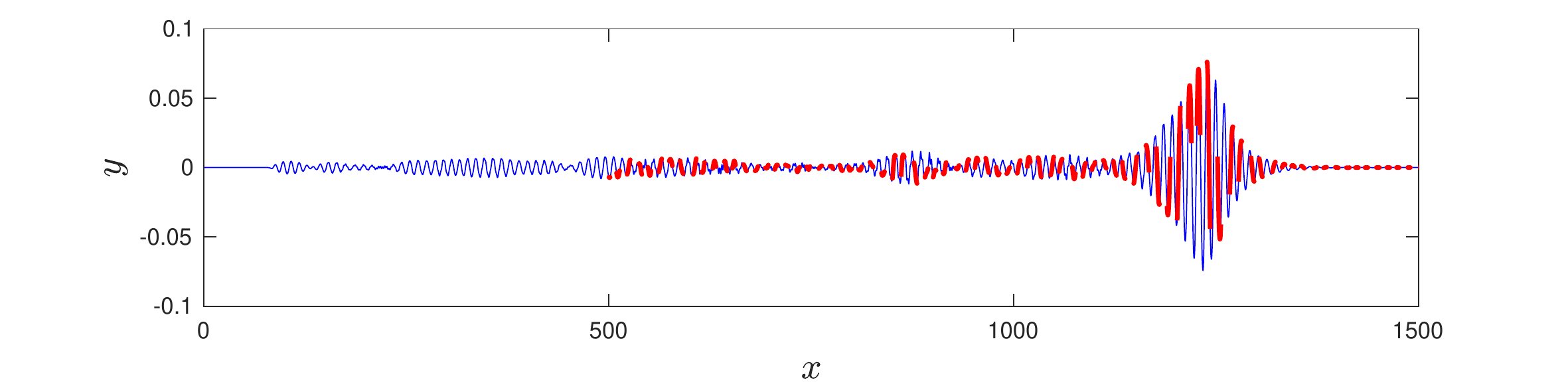}
\caption{$t = 1200$, $N_i = 124$, $L_i = 4$}
\end{subfigure}
\caption{Snapshots of $\eta$ at (a) $t = 0$, (b) $t = 300$, (c) $t = 750$, (d) $t = 1200$ for $(\varepsilon,a_0,k_0) = (0.05,0.08,0.6)$ 
and one realization of $(N_i,L_i) = (124,4)$.
Open water is represented in blue while ice floes are represented in red.}
\label{wave_a008_k06_xd2}
\end{figure}

To illustrate the scattering process and ensuing attenuation in the physical space for such incident waves and floe configurations,
Fig. \ref{wave_a008_k06_xd2} shows snapshots of $\eta$ at $t = 0$ (incident condition), $t = 300$ (soon after entering the ice cover),
$t = 750$ and $t = 1200$ (far into the ice cover) for $(a_0,k_0) = (0.08,0.6)$ in the weakly nonlinear regime $\varepsilon = 0.05$
and for one realization of $(N_i,L_i) = (124,4)$.
As indicated earlier, the incident pulse is initially located near the left edge of the ice cover and travels from left to right.
The computational domain being quite long, not its entire length is depicted here in order to show details
of the radiative tail that develops behind the wave group as it travels through the floe field.
The relative calmness observed near $x = 0$ is indicative of a sponge layer specified there, as mentioned in Sec. 3.1.
For graphical purposes, the individual floes are associated with the values of $f(x) \in [1 - \delta, 1 + \delta]$
(with $\delta = 10^{-3}$) to take floating-point arithmetic into account.
Note that, due to the large aspect ratio of these plots and the highly oscillatory nature of this incident condition, 
the sparse floe distribution is not so well rendered especially at $t > 0$ inside the main pulse where the short floes appear to be squeezed.
During propagation, the wave packet tends to gradually decrease in amplitude and expand in width 
while continually shedding a dispersive tail in the opposite direction.
This emission takes the form of a train of smaller wave packets, 
nonetheless the initial pulse retains a certain coherency and localized shape over the span of the ice field.

\begin{figure}[t!]
\centering
\begin{subfigure}[h]{1.1\textwidth}
\includegraphics[width=\linewidth]{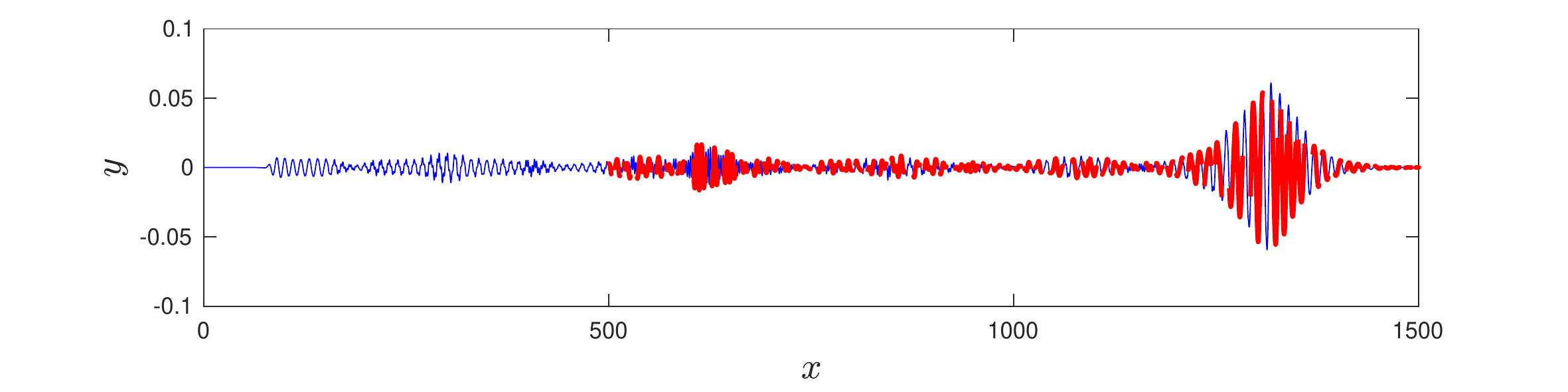}
\caption{$t = 1200$, $N_i = 124$, $L_i = 8$}
\end{subfigure}
\begin{subfigure}[h]{1.1\textwidth}
\includegraphics[width=\linewidth]{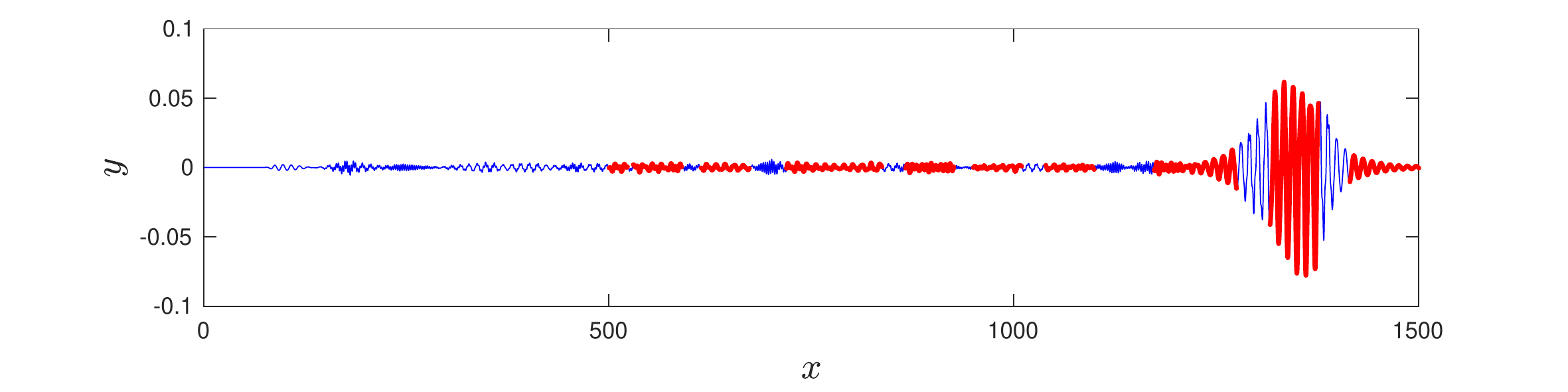}
\caption{$t = 1200$, $N_i = 20$, $L_i = 60$}
\end{subfigure}
\begin{subfigure}[h]{1.1\textwidth}
\includegraphics[width=\linewidth]{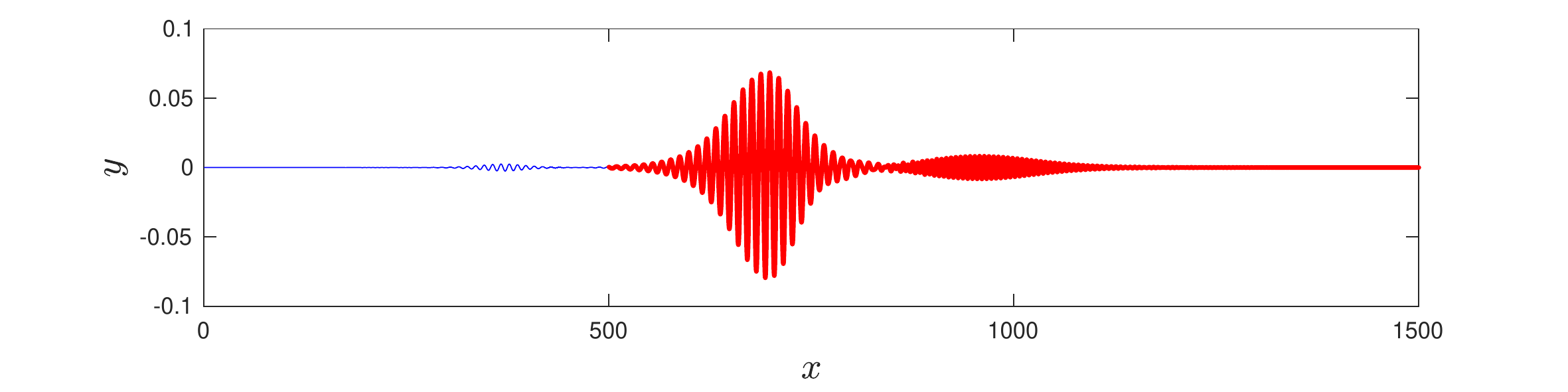}
\caption{$t = 360$, $N_i = 1$, $L_i = 1600$}
\end{subfigure}
\begin{subfigure}[h]{1.1\textwidth}
\includegraphics[width=\linewidth]{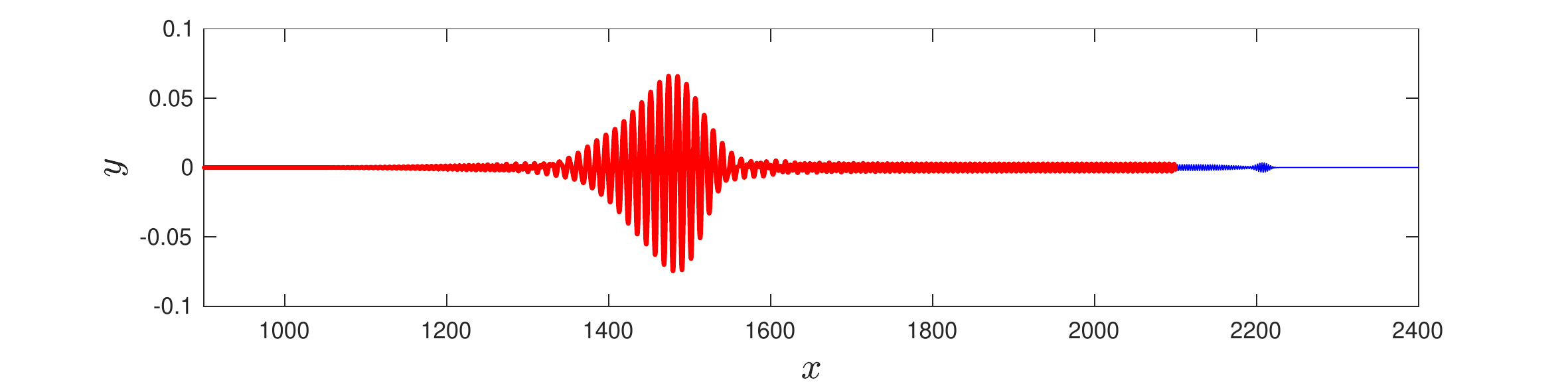}
\caption{$t = 1200$, $N_i = 1$, $L_i = 1600$}
\end{subfigure}
\caption{Snapshots of $\eta$ at $t = 1200$ for one realization of (a) $(N_i,L_i) = (124,8)$ and (b) $(N_i,L_i) = (20,60)$.
Snapshots of $\eta$ at (c) $t = 360$ and (d) $t = 1200$ for $(N_i,L_i) = (1,1600)$.
In all these cases, the wave regime is $(\varepsilon,a_0,k_0) = (0.05,0.08,0.6)$.
Open water is represented in blue while ice floes are represented in red.}
\label{wave_a008_k06}
\end{figure}

As a comparison, Fig. \ref{wave_a008_k06} presents the solution at $t = 1200$ for the same pair $(a_0,k_0) = (0.08,0.6)$
in the denser floe settings $(N_i,L_i) = (124,8)$ and $(20,60)$.
Again, only one realization of each of these configurations is portrayed.
The overall picture for $(N_i,L_i) = (124,8)$ is found to be similar to that for $(N_i,L_i) = (124,4)$.
However, notable differences can be discerned as compared to the larger-floe case $(N_i,L_i) = (20,60)$
where the radiative component is smaller in amplitude and the main pulse is more distorted.
This result is consistent with the expectation that multiple scattering should be important for high levels of ice fragmentation
and that scattering should be significant when the wavelength is comparable to the floe size.
Indeed, among these three configurations, the backscattering appears to be stronger in the intermediate case 
$(N_i,L_i) = (124,8)$ for which $\lambda_0 = 10 \sim L_i = 8$.
When $(N_i,L_i) = (20,60)$, the presence of larger floes tends to distort the passing wave packet further
because waves in sea ice and open water have different dispersion properties as stipulated by \eqref{lindisp}.

In Fig. \ref{wave_a008_k06}(c,d) for the special setting $(N_i,L_i) = (1,1600)$, we see that when 
the incident wave impinges on the ice cover, a small component is radiated forward and then travels ahead of the main pulse,
which is consistent with the linear prediction \eqref{speed} for the group speed being higher in pack ice than in open water.
A similar phenomenon was reported in \cite{gp17} for such a floe configuration when perturbed by solitary waves on shallow water.
Aside from this initial disturbance with a small wave packet being emitted and separating from the main pulse,
scattering is found to be minimal in this case.
The leading undulation broadens in the early stages but remains coherent and localized,
traveling faster than its counterparts in a more fragmented ice cover.
Such a solution bears a resemblance to long-lived nonlinear wave packets arising near the critical speed 
$c_{\min} = 2/3^{3/8} = 1.32$ (for $k_{\min} = 1/3^{1/4} = 0.76$) in homogeneous sea ice over deep water, 
as computed by Guyenne and P\u ar\u au \cite{gp12}.

\begin{figure}[t!]
\centering
\begin{subfigure}[h]{1.1\textwidth}
\includegraphics[width=\linewidth]{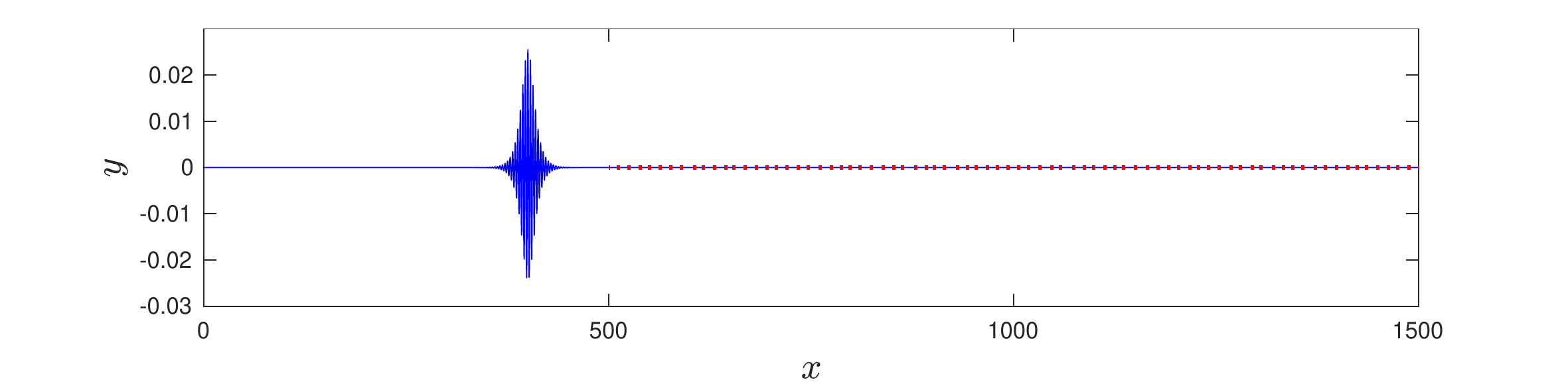}
\caption{$t = 0$, $N_i = 124$, $L_i = 4$}
\end{subfigure}
\begin{subfigure}[h]{1.1\textwidth}
\includegraphics[width=\linewidth]{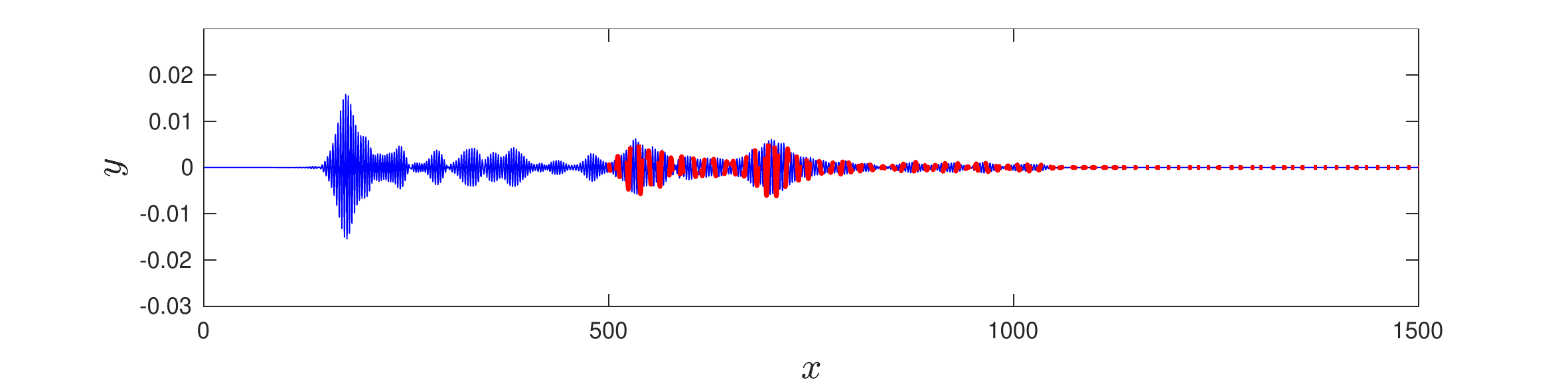}
\caption{$t = 1200$, $N_i = 124$, $L_i = 4$}
\end{subfigure}
\begin{subfigure}[h]{1.1\textwidth}
\includegraphics[width=\linewidth]{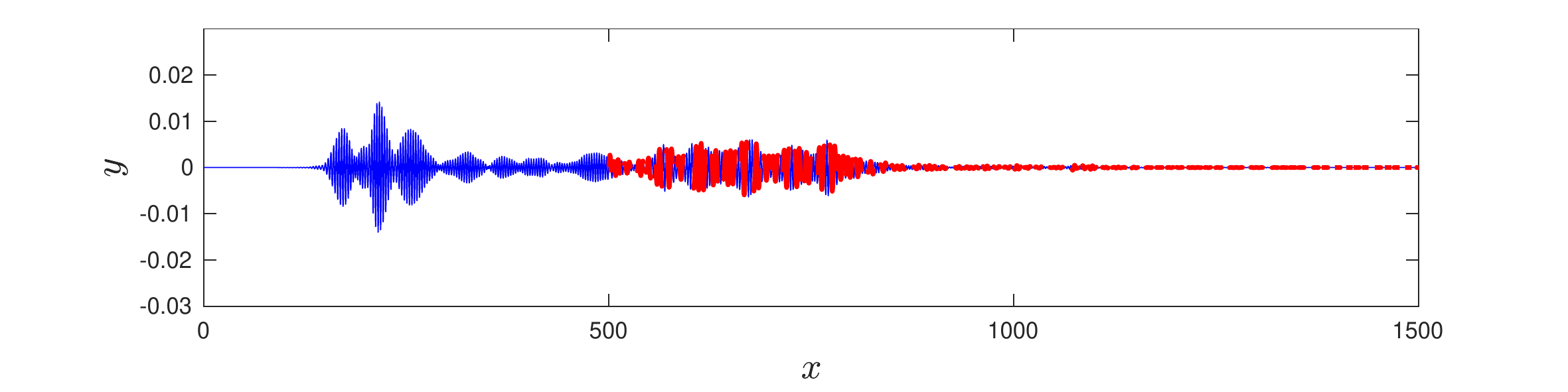}
\caption{$t = 1200$, $N_i = 124$, $L_i = 8$}
\end{subfigure}
\begin{subfigure}[h]{1.1\textwidth}
\includegraphics[width=\linewidth]{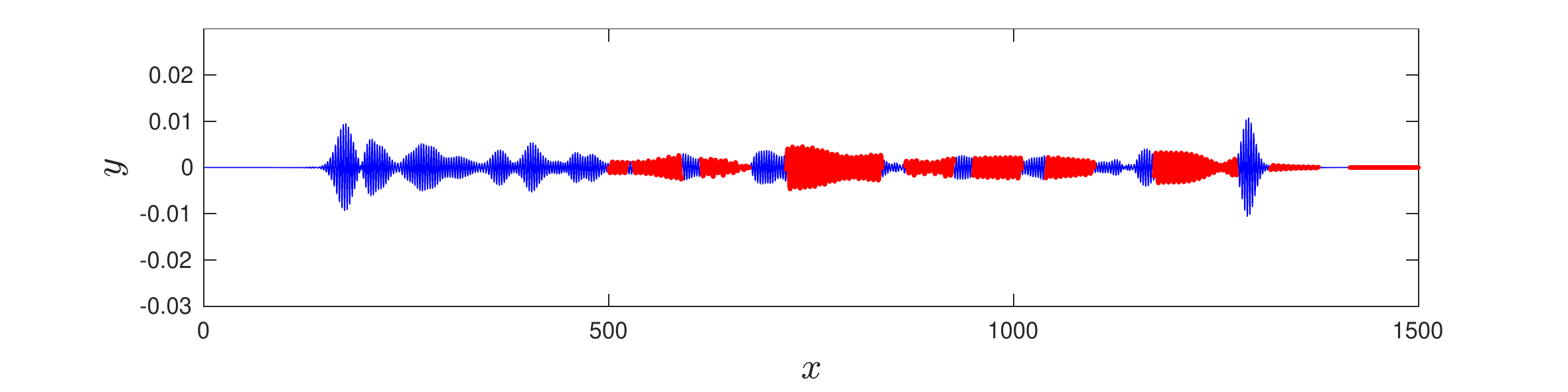}
\caption{$t = 1200$, $N_i = 20$, $L_i = 60$}
\end{subfigure}
\caption{Snapshots of $\eta$ at (a) $t = 0$ and (b) $t = 1200$ for one realization of $(N_i,L_i) = (124,4)$.
Snapshots of $\eta$ at $t = 1200$ for one realization of (c) $(N_i,L_i) = (124,8)$ and (d) $(20,60)$.
In all these cases, the wave regime is $(\varepsilon,a_0,k_0) = (0.05,0.02,2)$.
Open water is represented in blue while ice floes are represented in red.}
\label{wave_a002_k2}
\end{figure}

The larger the wavenumber $k_0$, the stronger the scattering, with a more prominent tail developing
behind the incident pulse. As a result, the latter quickly decays and does not go far into the ice field 
(see Fig. \ref{wave_a002_k2} for $\varepsilon = 0.05$ and $k_0 = 2$).
In this short-wave limit, scattering is so strong that the radiative tail displays a well-defined leading pulse
(moving backward) that is commensurate with the initial condition.
Over the range of parameter values considered, we observe that both regimes $\varepsilon = 0.05$ and $0.2$ 
exhibit similar qualitative features of wave propagation and scattering for the same values of $k_0$ and $(N_i,L_i)$.
This is in line with our previous comment that the initial solution tends to become less nonlinear due to the subsequent attenuation.
Indeed, for $\varepsilon = 0.2$, initially steep waves rapidly spread when interacting with sea ice.
Wave profiles in this case are thus not displayed for convenience,
but a more quantitative comparison between $\varepsilon = 0.05$ and $0.2$ will be provided in the next section 
when discussing estimates of the decay rate.

The present nonlinear simulations confirm that scattering is a major contributor to wave attenuation in broken floe fields
(when the wavelength is comparable to the floe size), which agrees with previous studies 
that have tested predictions from linear scattering theory against field measurements \cite{km08,wsgcm88}.
Moreover, our finding about the persistence of a localized progressive waveform that scatters little 
along an extended uniform floe (Fig. \ref{wave_a008_k06}c,d) is compatible with observations from Kohout et al. \cite{kwdm14} 
who reported that large waves in Antarctic sea ice decay more slowly than the expected exponential rate
and can persist hundreds of kilometers into the ice pack.
Our results for $(N_i,L_i) = (1,1600)$ also support accounts of an enduring (and even enhanced) group structure 
for surface waves in Arctic sea ice, as described by Gemmrich et al. \cite{gmt21} and Thomson et al. \cite{tgrca19}.

\subsection{Calculation of attenuation rates}

To assess more quantitatively wave attenuation for each set of $(N_i,L_i,a_0,k_0)$, 
we take the averaged values $\overline \eta$ of the simulated surface elevation over the ten realizations of $(N_i,L_i)$.
The goal of this averaging process is twofold:
it helps remove possibly non-generic realization-dependent effects and helps minimize fluctuations
across the heterogeneous medium in anticipation of the next fitting step to estimate the spatial attenuation rate.
More specifically, for each grid point in the ice field ($500 \le x \le 2100$), we evaluate the $L^2$ norm
\begin{equation} \label{ener}
S(x) = \frac{1}{T_2 - T_1} \int_{T_1}^{T_2} \big( \overline \eta(x,t) \big)^2 dt \,,
\end{equation}
which is analogous to the wave energy spectrum as measured by an array of sensors in field experiments.
To provide a uniform calculation of \eqref{ener} across the board, the scanning time interval is given by 
$[T_1,T_2] = [200,1200]$ for all the cases considered.
It is selected as a compromise for the observation window to be sufficiently long, while ensuring that the wave group has entered 
the ice field after $t = T_1$ and that possible undesirable wave reflection from the domain boundaries is negligible prior to $t = T_2$.
Indeed, despite our efforts, small wave reflection from the sponge layers was inevitably produced 
after an extended lapse of time, especially during the propagation of broad or long wave groups.
The trapezoidal rule is applied to compute \eqref{ener} with sampling step size $1500 \Delta t = 3$.
Inspired by linear theory, the spatial attenuation rate $\alpha$ of the incident wave energy
can be estimated by least-squares fitting an exponential function of the form
\begin{equation} \label{fit}
S(x) \sim e^{-\alpha x} \,,
\end{equation}
to computations of \eqref{ener} over a portion of the region $500 < x < 2100$ occupied by the ice field.

\begin{figure}[t!]
\centering
\begin{subfigure}[h]{0.49\textwidth}
\includegraphics[width=\textwidth]{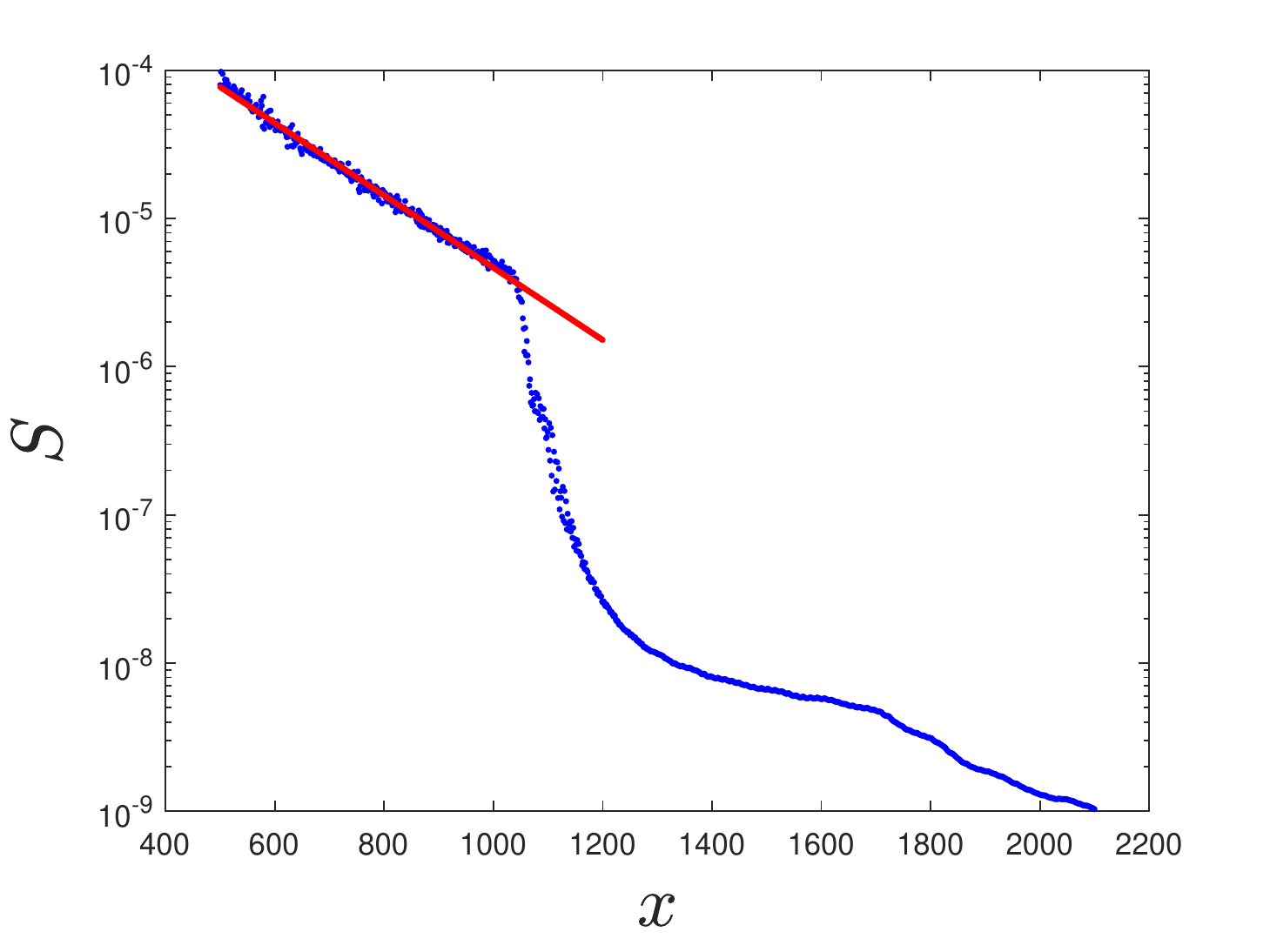}
\caption{}
\end{subfigure}
\hfill
\begin{subfigure}[h]{0.49\textwidth}
\includegraphics[width=\textwidth]{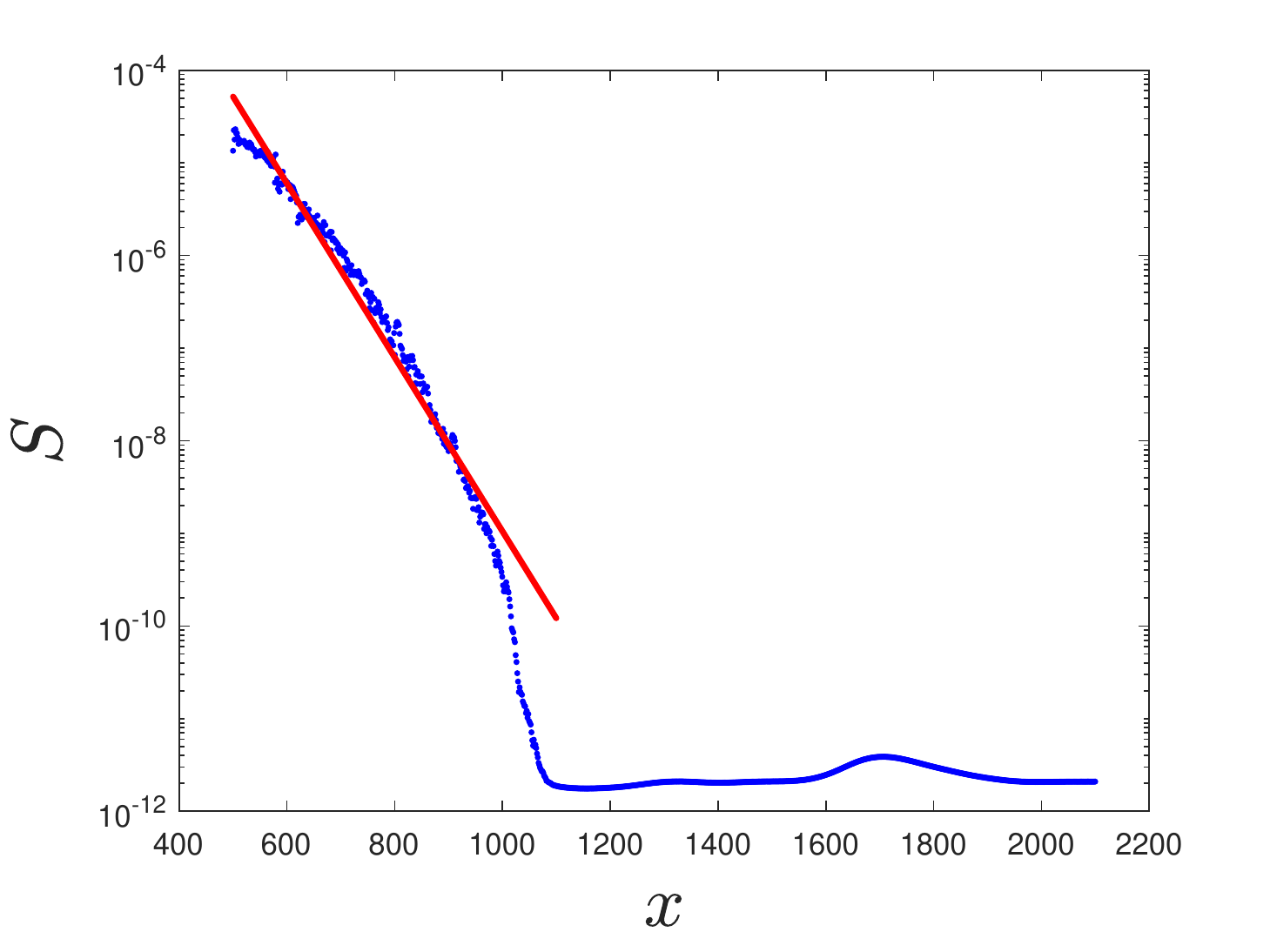}
\caption{}
\end{subfigure}
\caption{Variation of $S(x)$ across the ice field (blue dots) and corresponding exponential fit (red line)
for (a) $(\varepsilon,a_0,k_0) = (0.2,0.11,1.8)$ and (b) $(\varepsilon,a_0,k_0) = (0.05,0.02,2.2)$.
In both cases, the floe configuration is $(N_i,L_i) = (124,4)$.}
\label{spectra_fit}
\end{figure}

As an illustration, Fig. \ref{spectra_fit} shows fits to \eqref{ener} for $(a_0,k_0,N_i,L_i) = (0.02,2.2,124,4)$ and $(0.11,1.8,124,4)$
where $S(x)$ is plotted in semilog units for clarity so that the exponential dependence \eqref{fit} appears as a straight line.
Despite averaging $\eta$ over the ten realizations of $(N_i,L_i)$ and integrating \eqref{ener}
over the time interval $[T_1,T_2]$, small fluctuations still occur but nonetheless the profile of $S(x)$ can be clearly identified.
In these two cases, the exponential fit is applied to the window $500 < x < 1000$ beyond which
wave effects are found to be even weaker by orders of magnitude, consistent with the fast wave decay as noted above for large $k_0$.
Such plots confirm that it is reasonable to assume \eqref{fit} in order to obtain a measure of the attenuation rate for incident wave packets,
even though our computations based on \eqref{eq-eta}--\eqref{eq-xi} are nonlinear.
Similar graphs arose for other sets of parameter values but the fit was taken over different windows
(confined to $500 < x < 2100$) where an exponential-like behavior was detected.
By repeating this procedure for a range of $k_0$, 
a set of values for $\alpha$ as a function of wave frequency can be collected for each triplet $(\varepsilon,N_i,L_i)$.
There are some similarities with the procedure typically adopted in field experiments on this subject:
measurements are filtered in some way to help remove noise and an exponential functional dependence is assumed 
to extract an apparent decay rate from directional wave spectra \cite{cheng17,mbk14,wsgcm88}.
We also remark that our focus here is on determining the spatial attenuation rate $\alpha$ for wave groups,
which differs from the estimation of temporal attenuation rates for solitary waves as conducted in \cite{gp17}.
For that particular problem, no reference results were available to compare with
and, moreover, only a few different incident wave heights were examined in that previous study.

Figure \ref{compare_all} depicts the spatial attenuation rate $\alpha$ as a function of incident wave frequency $f$
for all eight cases of $(\varepsilon,N_i,L_i)$ involved.
Considering the weakly nonlinear nature of the incident conditions as prescribed here,
the corresponding wave frequency is approximated by $f = \omega/(2\pi)$ where $\omega = \omega_0  \sqrt{1 + \varepsilon^2}$
according to Stokes theory.
Overall, these results on decay rate support our previous observations based on the description of wave profiles.
There is a general tendency for $\alpha$ to increase with $f$,
which is consistent with typical predictions from linear (scattering or viscoelastic) theory.
The sparser the floe field or the lower the incident wave steepness, the faster the increase of $\alpha$ with $f$.
A sharp rise occurs for $(N_i,L_i) = (124,4)$ and $(124,8)$ in the high-frequency region $f > 0.19$,
which corresponds to the wavenumber range $k_0 > 1.4$ where scattering is strong because the wavelength is comparable to the floe size.
The largest value of $\alpha$ that we calculated is attained at $f = 0.24$ (i.e. $k_0 = 2.2$) 
for $\varepsilon = 0.05$ and $(N_i,L_i) = (124,4)$, which indeed satisfies $\lambda_0 = 2.8 \sim L_i = 4$. 
Such waves of small amplitude and short wavelength tend to quickly disintegrate as they propagate into a broken ice field.
This process is aggravated by the high level of ice fragmentation ($N_i = 124$) in these two floe settings, 
which promotes multiple wave scattering.

On the other hand, for $(N_i,L_i) = (20,60)$ with larger floes, the decay rate remains small and does not vary much
across the spectrum of wave frequencies being probed.
In this case, the attenuation curve for $\varepsilon = 0.05$ closely resembles that for $\varepsilon = 0.2$.
This uniformity and similarity in $\alpha$ between $\varepsilon = 0.05$ and $0.2$ is even more pronounced
in the single-floe configuration $(N_i,L_i) = (1,1600)$.
The presence of ice modifies the dispersion properties of incident waves and thus affects their speed and shape,
but coherent structures persist throughout the homogeneous ice cover without experiencing much attenuation, 
as evidenced by Fig. \ref{wave_a008_k06}(d).

Interestingly, we notice that the rise of decay rate is not strictly monotonic as a function of wave frequency in some situations.
There seems to be a hump in the graph of $\alpha$ around $f = 0.16$ (i.e. $k_0 = 1$),
which is most apparent for highly broken floe fields ($N_i = 124$, $L_i = \{ 4, 8 \}$)
and whose amplitude is a bit larger for steeper incident waves ($\varepsilon = 0.2$).
This finding is reminiscent of the roll-over effect that has been reported in field measurements \cite{wsgcm88}.
While various possible causes have been suggested, this intriguing phenomenon is still not well understood
and has generally eluded linear scattering or viscoelastic models.
We defer a more detailed discussion to the next section when comparing our numerical estimates to field data.
To our knowledge, this is the first time that such a roll-over has been revealed from phase-resolved nonlinear simulations
of wave attenuation in sea ice.

\subsection{Comparison with field observations}

To provide some assessment about the physical relevance of our numerical results, we now compare them to a selection of field measurements.
We choose to test against field experiments rather than laboratory experiments
because scattering is believed to be the dominant mechanism for wave attenuation in broken ice covers
and thus is more likely to be observed in the former situation.
Furthermore, our choice of non-dimensionalization as described in Sec. 2.2 and our choice of incident wavenumbers as discussed in Sec. 3.2
would be more suitable for the oceanic environment.
Because we originally did not attempt to set up our simulations under specific conditions corresponding to any of these field experiments,
our expectation at best would be to obtain some qualitative agreement.
Moreover, as opposed to calibration studies for linear viscoelastic-type models \cite{cheng17,desanti18,mms15,xg22},
we do not attempt here to accurately fit our numerical predictions on $\alpha$ to the experimental data 
by minimizing any objective function over a range of possible values for rheological parameters.
Such a procedure in the context of direct simulations of nonlinear time-dependent partial differential equations would be quite challenging to implement.
Instead, we fix the rheological parameters of the ice cover (and their dimensionless form) by using representative values as stated in Sec. 2.1
and, for each set of field data from this selection, we present it together with a set of numerical estimates that is deemed to be its closest match
among the eight cases of $(\varepsilon,N_i,L_i)$ we simulated.
A visual assessment was sufficient to find matching results in our comparative study.

These field measurements are on the spatial attenuation rate of the wave energy spectrum in sea ice,
which is consistent with the $L^2$ norm \eqref{ener} that we adopt to quantify wave attenuation from our computations.
Motivated by the discussion in the previous section, our selection of experimental data includes observations by Wadhams et al. \cite{wsgcm88}
from the Greenland and Bering Seas in the late 1970s and early 1980s.
A series of experiments were conducted where wave attenuation was measured along a line of stations 
extending from the open sea deep into an ice field.
At each station, a wave buoy was placed between floes to measure the local wave spectrum.
Among the measurements reported in \cite{wsgcm88} (see their Table 2), 
we will select those from the Greenland Sea in 1979 and from the Bering Sea in 1983.
Other data sets (e.g. 1978 Greenland Sea or 1979 Bering Sea) are considered unreliable due to possibly larger experimental error
or undesired physical effects such as wave reflection from the fjords, as noted in \cite{km08}.

An intriguing aspect of the 1979 Greenland Sea and 1983 Bering Sea measurements is that they reveal a roll-over of decay rate
as a function of wave frequency, rather than a monotonic behavior.
Among past theoretical and numerical studies, we are aware of only a few that were able to reproduce such a phenomenon to some extent.
In the context of linear theory, the three-layer viscoelastic model of Zhao and Shen \cite{zs18}
predicts a roll-over that becomes more pronounced as the thickness of the middle turbulent layer increases.
However, no comparison with field data involving the roll-over was presented in their study.
A similar observation was made by Liu et al. \cite{lhv91} based on a linear model for a thin elastic plate lying over a fluid with eddy viscosity.
These authors derived a temporal rate of wave attenuation and divided it by the group speed to obtain a spatial rate.
Quoting \cite{lkdwgs17}, because this temporal rate is a monotonic function of frequency, the fact that the converted spatial rate 
exhibits roll-over may be due to the group speed being non-monotonic as indicated by \eqref{speed}.
Therefore, it is unclear from Liu et al.'s results whether the roll-over effect is an intrinsic feature of their thin-plate viscoelastic model
or is simply an artifact of their observational procedure.
Based on a linear analysis for potential flow, Olla et al. \cite{odd21} argued that inhomogeneities in the ice cover at scales
comparable to the wavelength significantly increase diffusion, producing a contribution to wave attenuation similar to 
what is observed in the ocean and usually explained by viscous effects.
The resulting attenuation spectrum is characterized by a peak at the scale of these inhomogeneities,
which could explain the roll-over at short wavelengths as reported by field experiments.
Recent assessment of a porous viscoelastic model for linear gravity waves in ice-covered seas \cite{xg22},
which was proposed earlier by Chen et al. \cite{cgg19}, demonstrates that it can emulate to some degree the roll-over from Arctic MIZ measurements.
In this poroelastic formulation, the non-monotonic behavior of decay rate is attributed to friction
induced by the relative motion between fluid and solid constituents of the ice cover.

As far as nonlinear models are concerned, a similar phenomenon has been observed in numerical simulations 
by Perrie and Hu \cite{ph96} and Li et al. \cite{lkdwgs17}.
Their results suggest that both wind forcing and nonlinear interactions must be active for a roll-over to develop.
We remark however that, unlike the present study, these authors used approximate phase-averaged wave forecasting models
(e.g. second-generation wave model, WAVEWATCH III).
To represent effects from wave-ice interaction, Perrie and Hu \cite{ph96} incorporated a scattering module 
where the distribution of ice floes is chosen among grid points of the computational domain,
while Li et al. \cite{lkdwgs17} employed effective parameterizations according to continuum viscoelastic theories.

For the purposes of the comparison with field measurements, our numerical results are converted back to dimensional estimates
based on the characteristic values for the length and time scales introduced in Sec. 2.2
(again we will retain the same notations for convenience).
Accordingly, the three different floe sizes $L_i = \{ 4, 8, 60 \}$ (in dimensionless units) become 
$L_i = \{ 61.2, 122.3, 917.3 \}$ m (in dimensional units).
The total length of the ice field now reads $L_c = 24.5$ km, with resolution $\Delta x = 2.3$ m.
Incidentally, the largest incident wavenumber that we prescribed is given by $k_0 = 2.2$ (in dimensionless units)
or $k_0 = 0.14$ m$^{-1}$ (in dimensional units), which corresponds to our smallest incident wavelength $\lambda_0 = 43.7$ m. 
The thin-plate approximation is thus well justified here since this value of $\lambda_0$ 
is significantly larger than the characteristic ice thickness $h = 1$ m.
Figure \ref{greenland} plots data on $\alpha$ versus $f$ from the 10 September 1979 Greenland Sea experiment 
together with our closest predictions corresponding to $(\varepsilon,N_i,L_i) = (0.2,124,61.2 \mbox{ m})$.
Note that, during this expedition, the ice cover was sparse and the floes were generally large.
The ice concentration was estimated to be ${\cal C} = 0.17$ from photograph analysis, 
with the floe size around ${\cal L}_i = 50$--$80$ m, as reported in \cite{km08,wsgcm88}.
These ice conditions turn out to be comparable to those in our best-matching simulations ($C = 0.31$, $L_i = 61.2$ m).
For clarity, the calligraphic style is employed to distinguish the measured values of ice concentration and floe size from the simulated values.

We see in Fig. \ref{greenland} that the range of measured frequencies falls within the simulated range 
(the latter being slightly larger than the former), and their respective resolutions are similar.
Overall, the agreement is found to be quite satisfactory, with the roll-over around $f = 0.13$ Hz 
and the decrease at low frequencies being reasonably well captured.
While the roll-over amplitude is a bit overestimated, its location and shape (spanning several data points) are well resolved by our numerical results.
For $f  > 0.15$ Hz (past the hump) where experimental data are not available, a sudden increase is predicted,
with $\alpha$ reaching its highest point at the end of the computed spectrum.
This extremal value turns out to be higher than the roll-over peak.
Therefore, it seems that the roll-over is not directly related to strong scattering induced by close wave resonance with the floes,
which rather occurs at larger frequencies (e.g. it can be checked that $\lambda_0 = 48 \mbox{ m} \sim L_i = 61 \mbox{ m}$ for $f = 0.18$ Hz).
By contrast, the roll-over around $f = 0.13$ Hz corresponds to a longer wavelength $\lambda_0 = 96$ m.
It is likely not connected to the critical wavenumber $k_{\min}$ at which the group and phase speeds coincide, 
because the associated wavelength $\lambda_{\min} = 126.4$ m is fairly different.
Moreover, this prediction on $k_{\min}$ is meant for waves propagating in homogeneous sea ice 
and thus does not refer to any specific floe size.

\begin{figure}[t!]
\centering
\begin{subfigure}{0.49\textwidth}
\includegraphics[width=\linewidth]{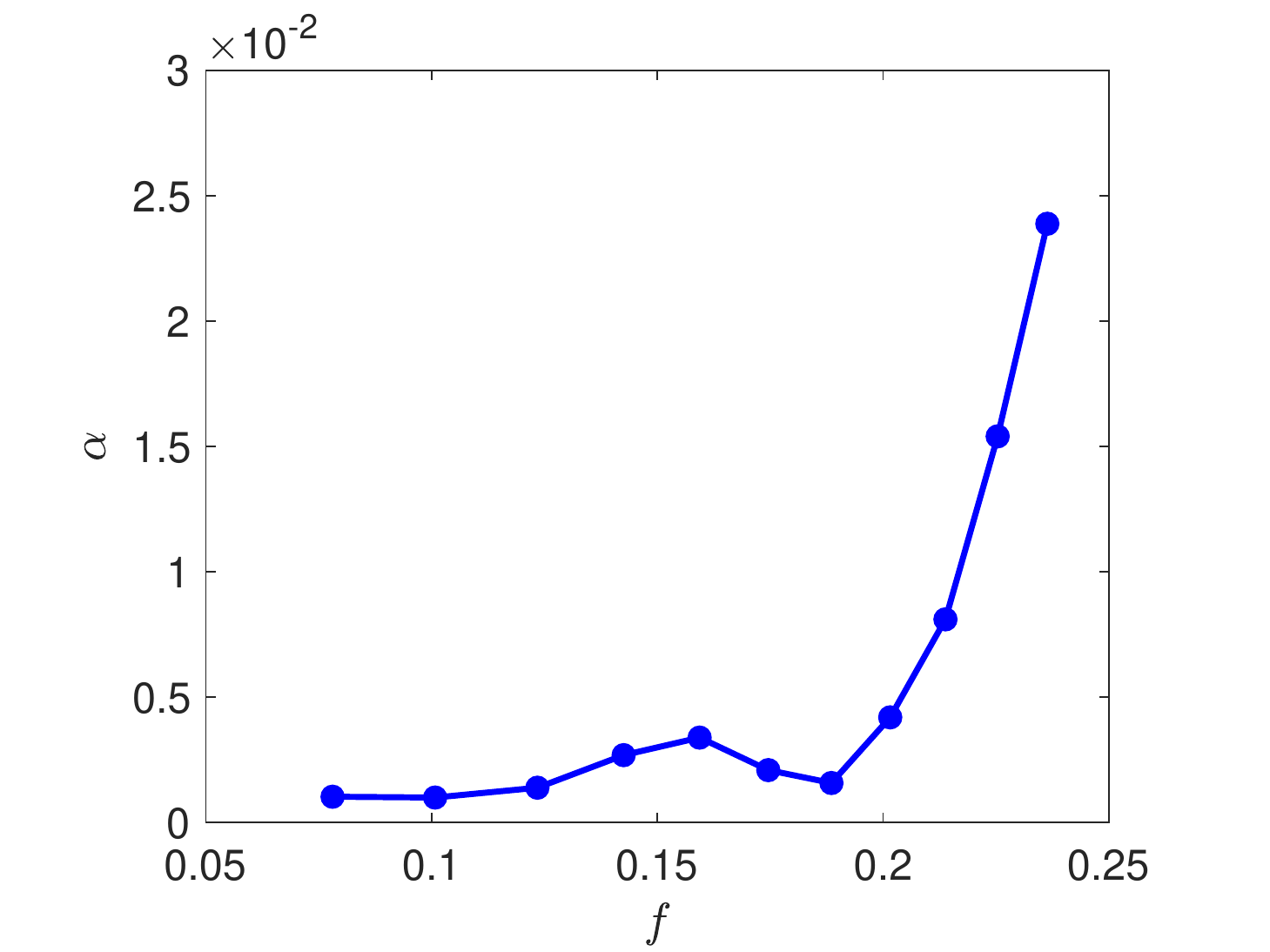}
\end{subfigure} 
\hfill
\begin{subfigure}{0.49\textwidth}
\includegraphics[width=\linewidth]{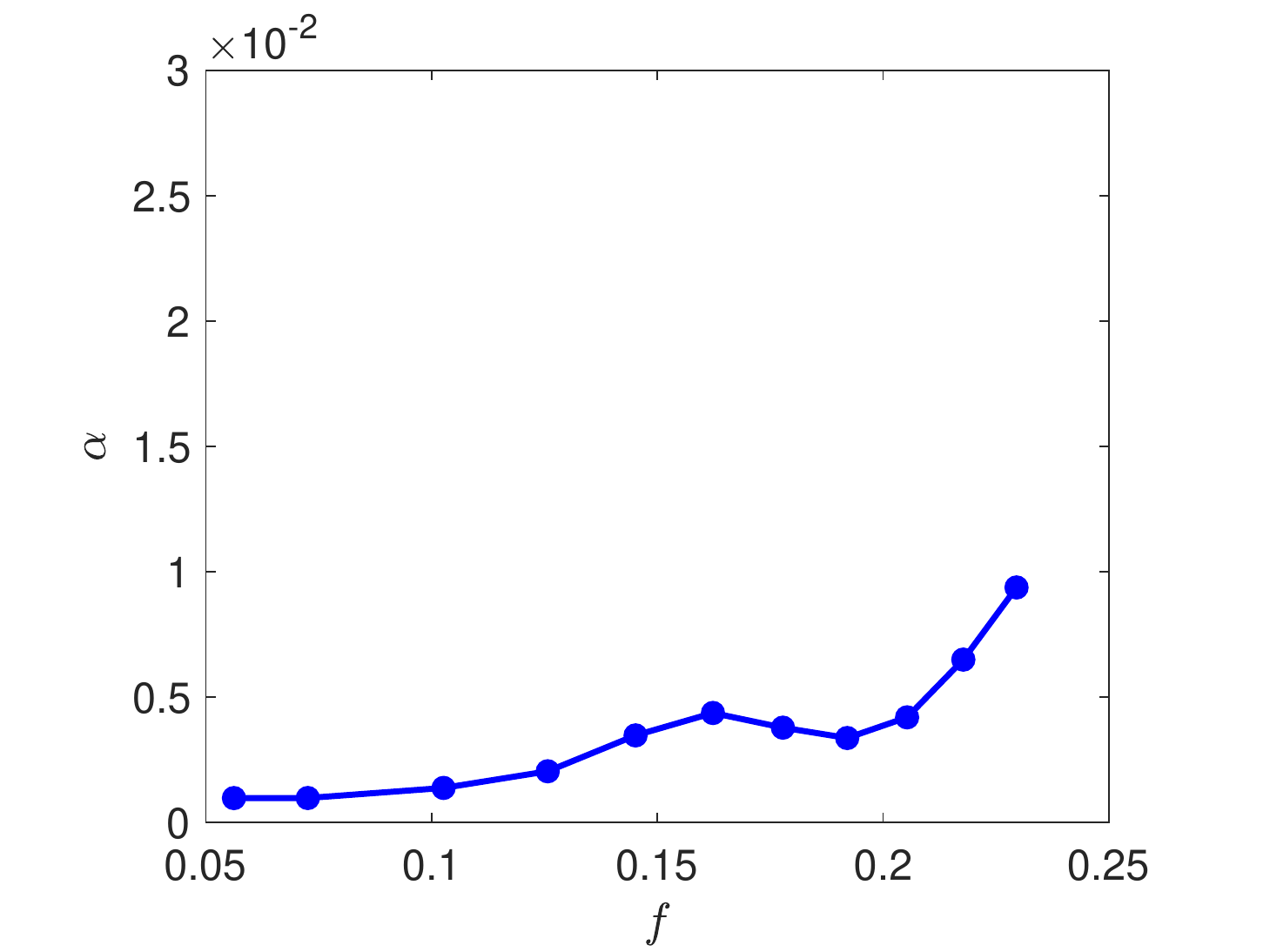}
\end{subfigure}
\begin{subfigure}{0.49\textwidth}
\includegraphics[width=\linewidth]{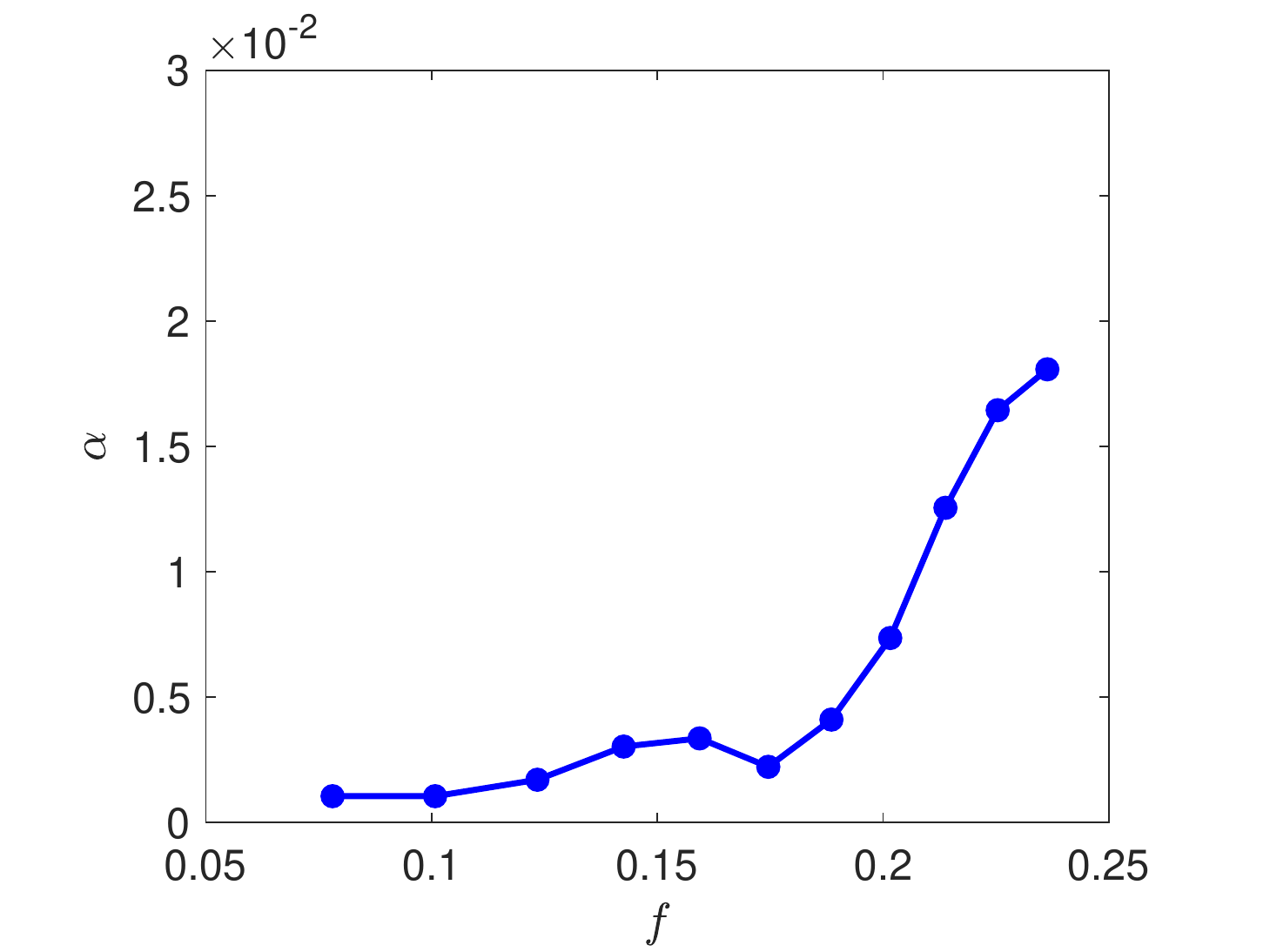}
\end{subfigure} 
\hfill
\begin{subfigure}{0.49\textwidth}
\includegraphics[width=\linewidth]{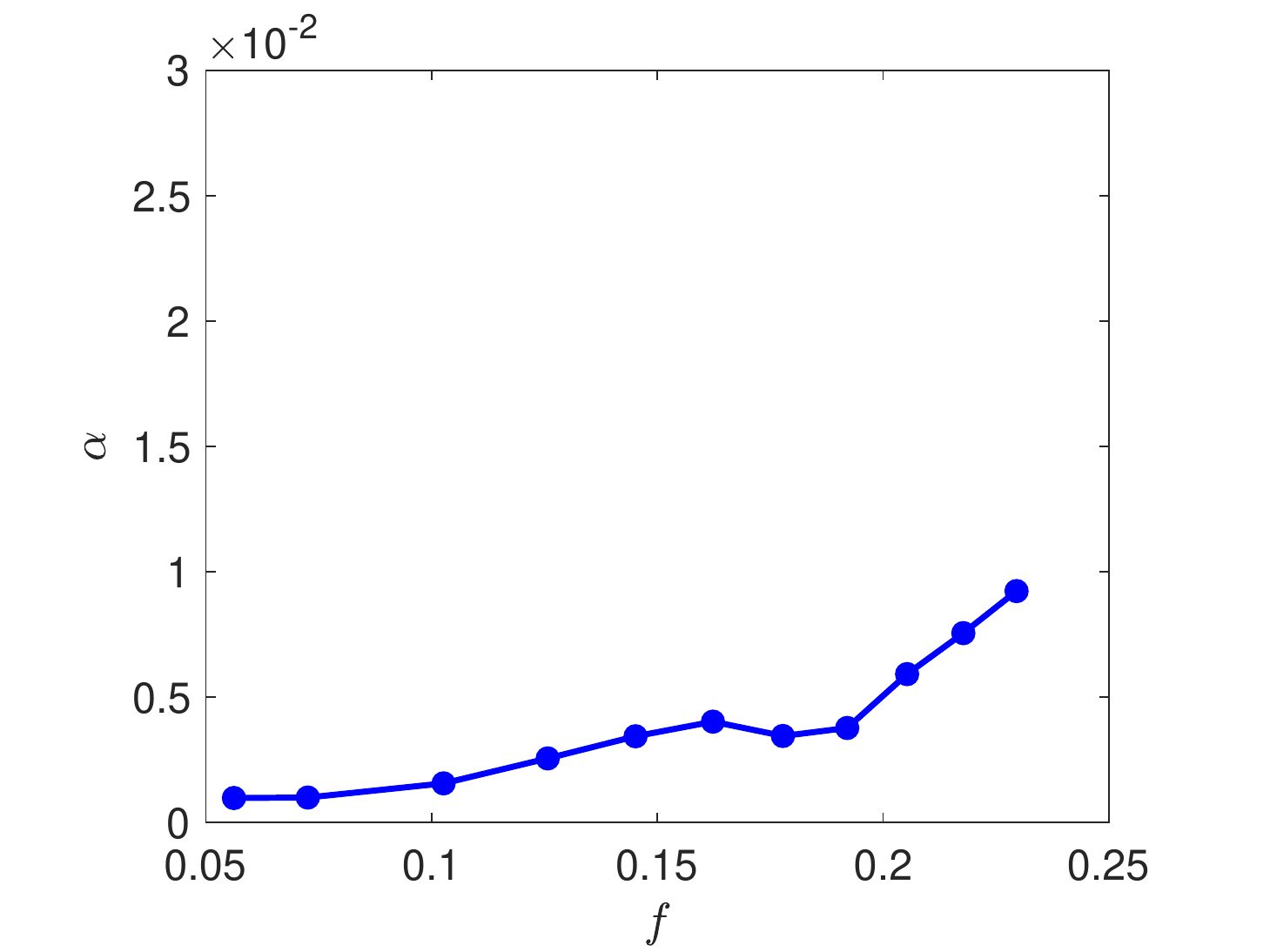}
\end{subfigure}
\begin{subfigure}{0.49\textwidth}
\includegraphics[width=\linewidth]{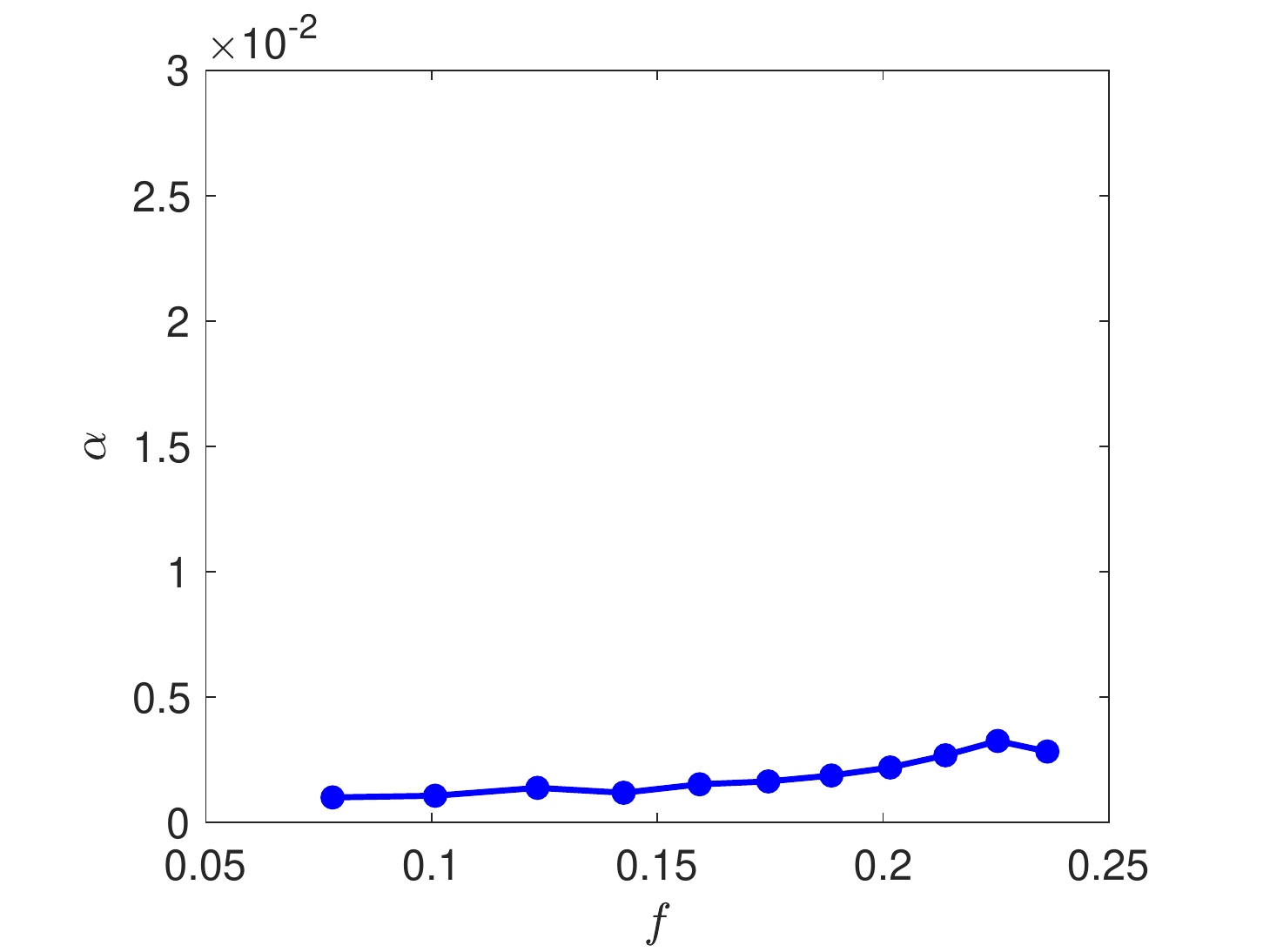}
\end{subfigure} 
\hfill
\begin{subfigure}{0.49\textwidth}
\includegraphics[width=\linewidth]{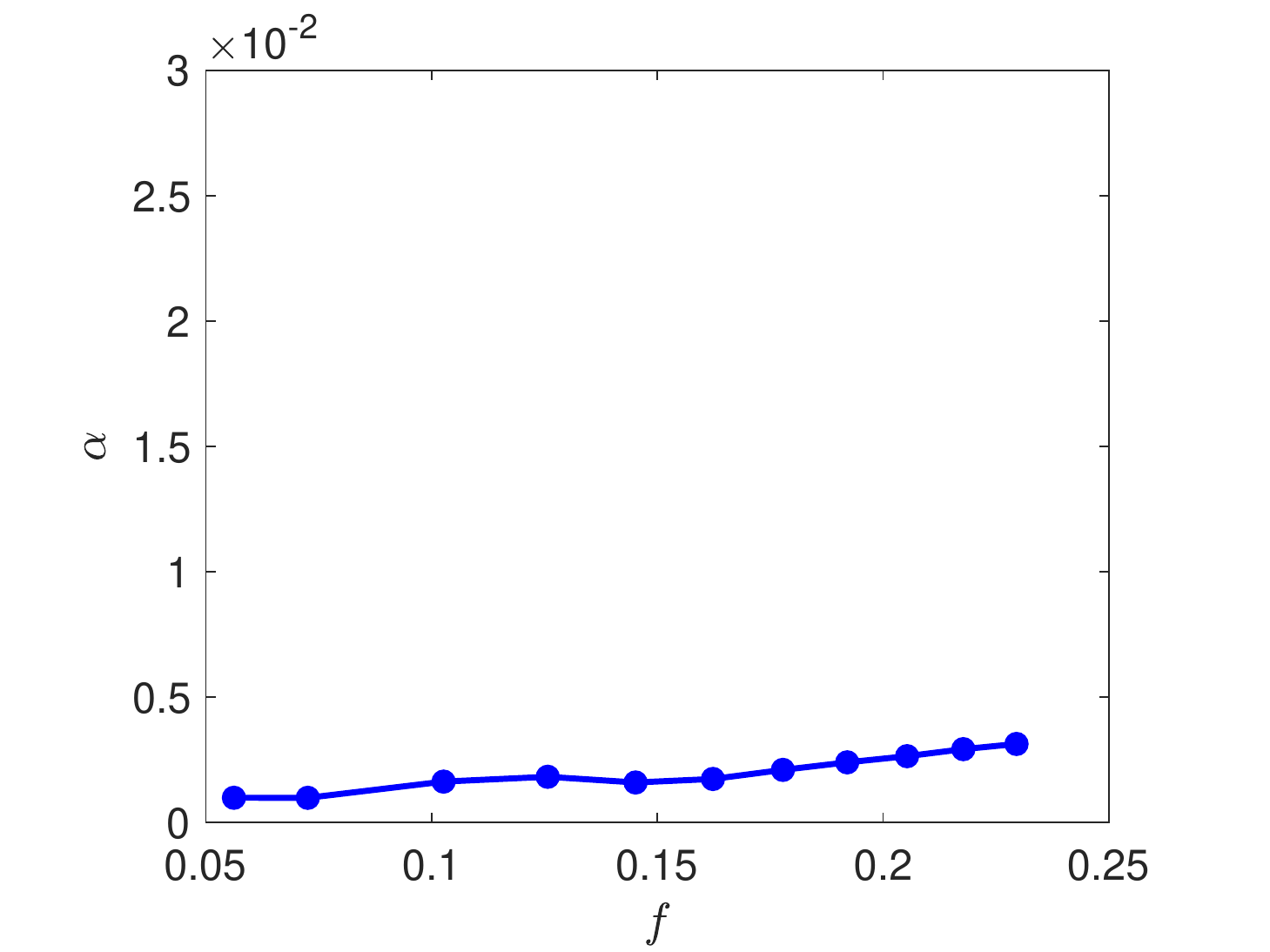}
\end{subfigure}
\begin{subfigure}{0.49\textwidth}
\includegraphics[width=\linewidth]{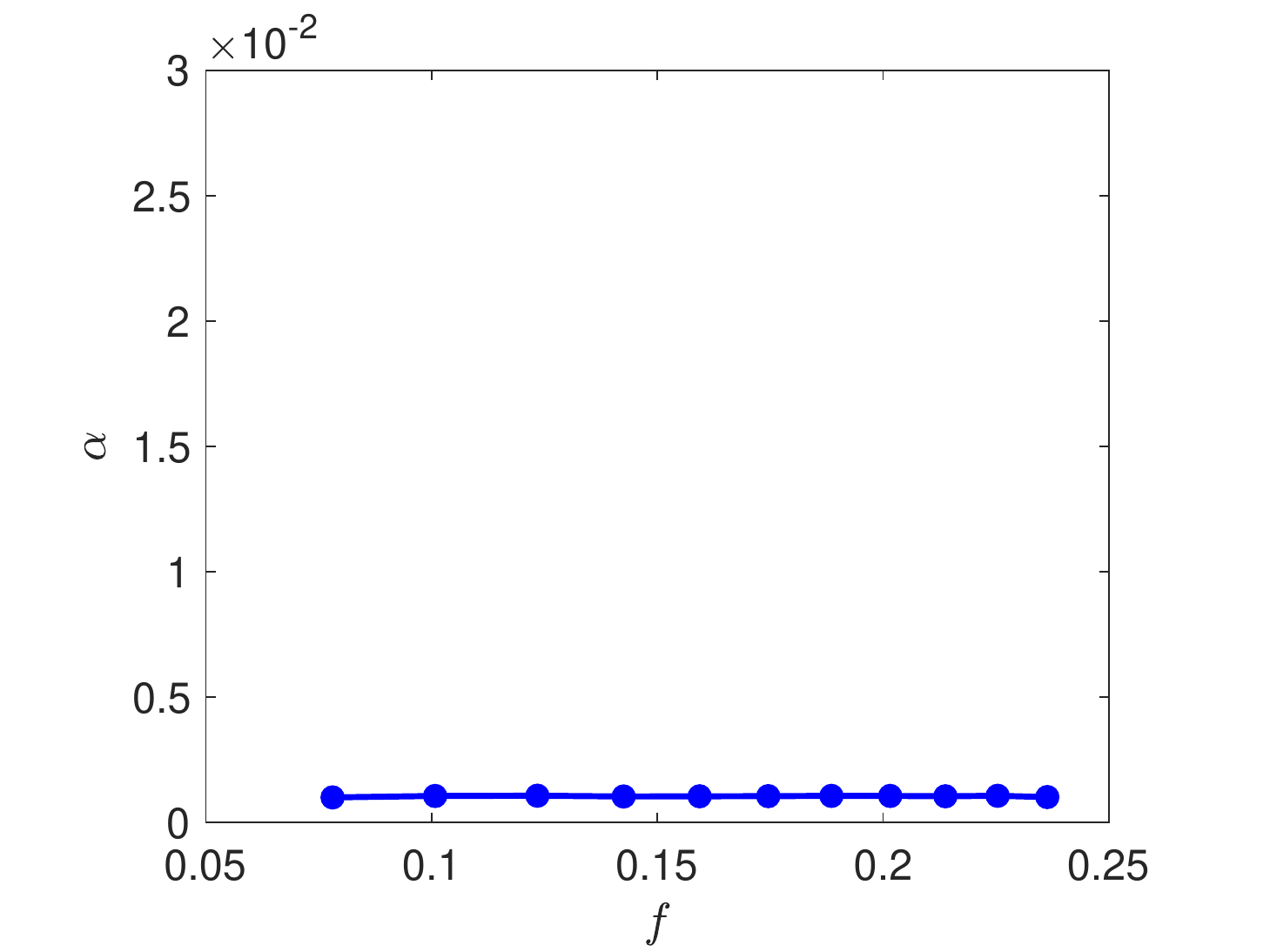}
\end{subfigure} 
\hfill
\begin{subfigure}{0.49\textwidth}
\includegraphics[width=\linewidth]{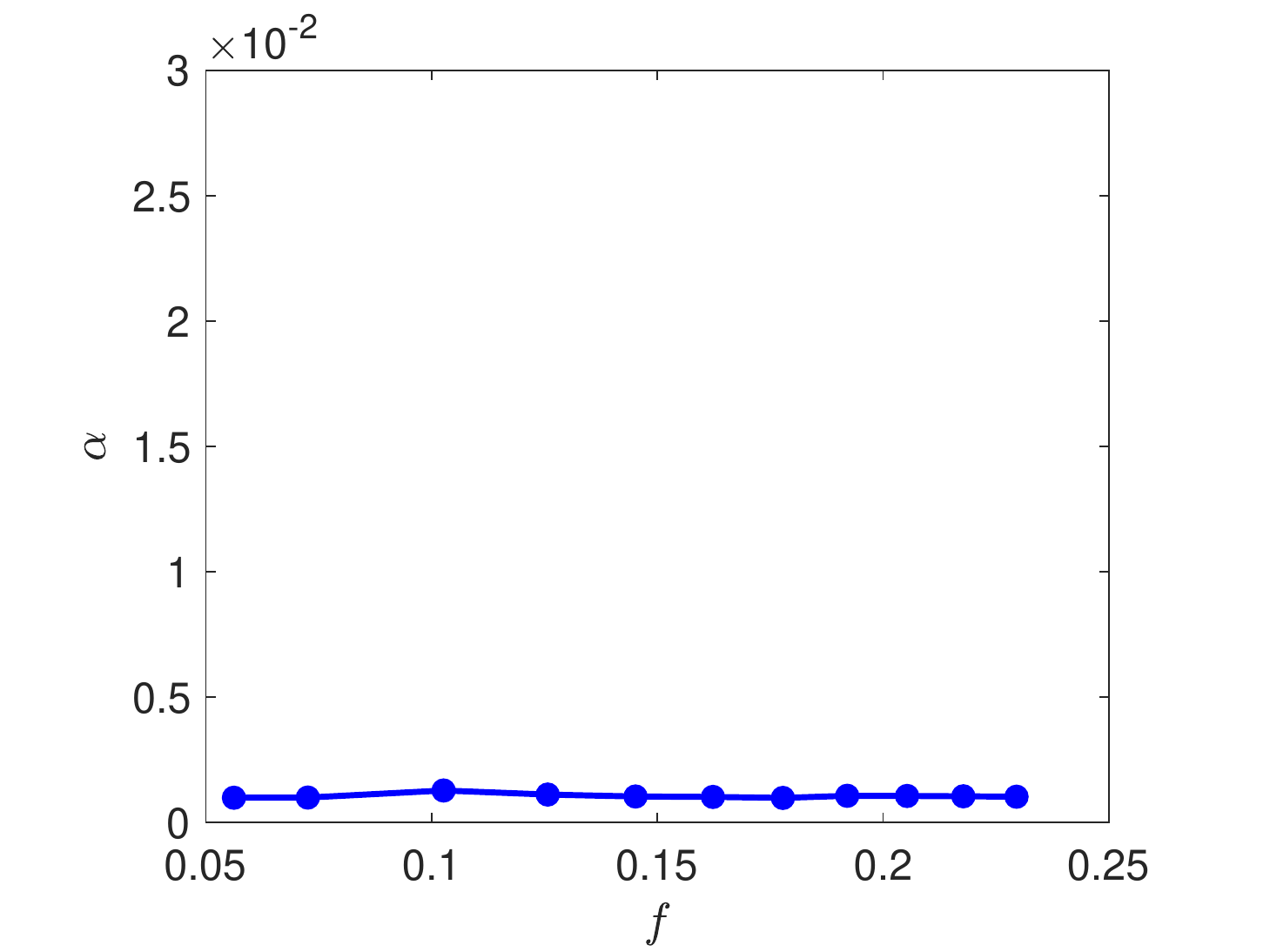}
\end{subfigure}
\caption{Attenuation rate vs. frequency for $\varepsilon = 0.05$ (left column) and $\varepsilon = 0.2$ (right column).
Rows from top to bottom: $(N_i,L_i) = (124,4)$, $(124,8)$, $(20,60)$, $(1,1600)$.}
\label{compare_all}
\end{figure}

\begin{figure}[t!]
\includegraphics[width=\linewidth]{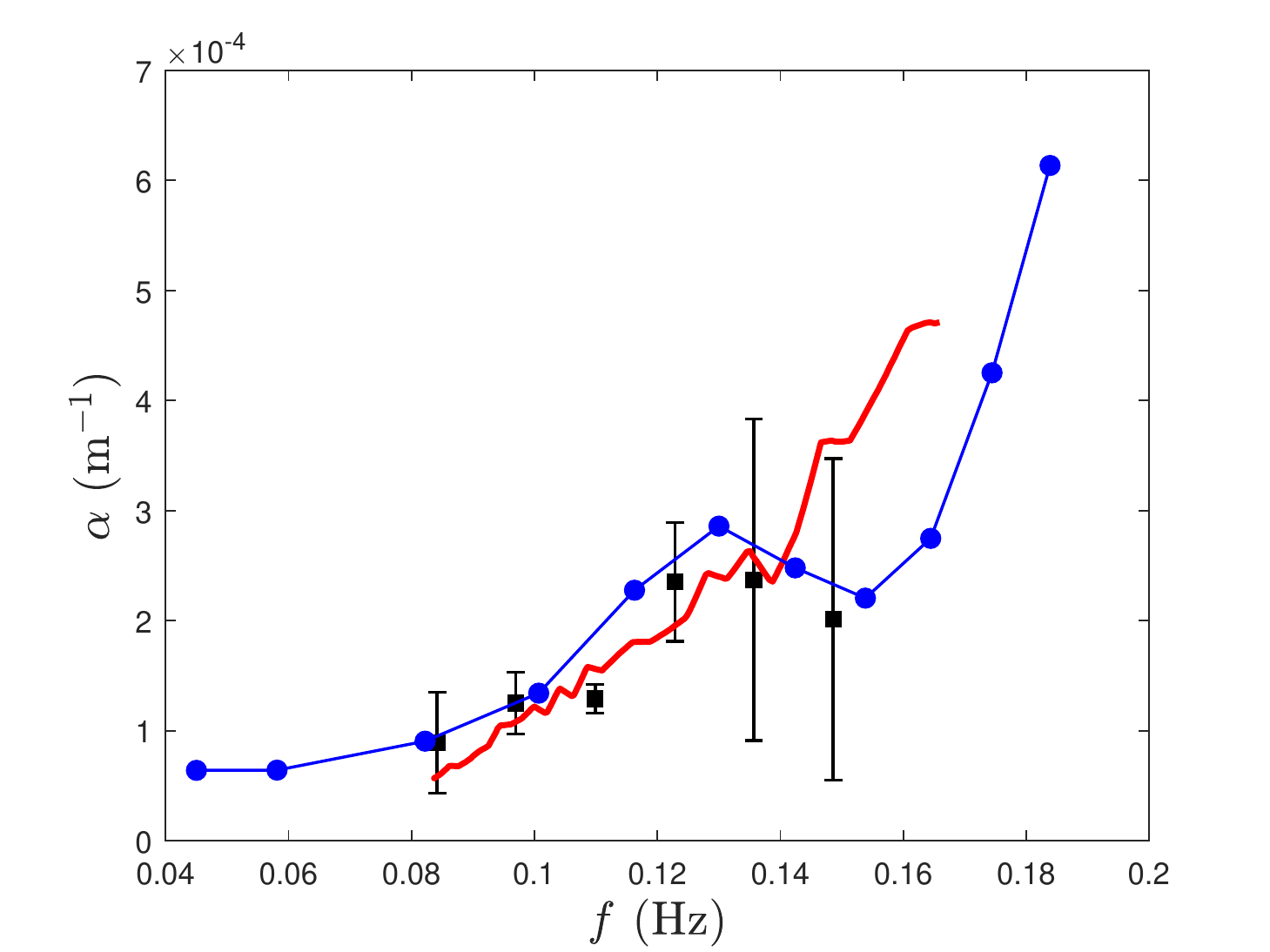}
\caption{Comparison of attenuation rate vs. frequency between measurements (black squares) 
from the 10 September 1979 Greenland Sea experiment \cite{wsgcm88} and our closest estimates (blue dots)
for $(\varepsilon,N_i,L_i) = (0.2,124,61.2 \mbox{ m})$.
Predictions (red line) from the linear scattering model of Kohout and Meylan \cite{km08} are also presented.}
\label{greenland}
\end{figure}

\begin{figure}[t!]
\includegraphics[width=\linewidth]{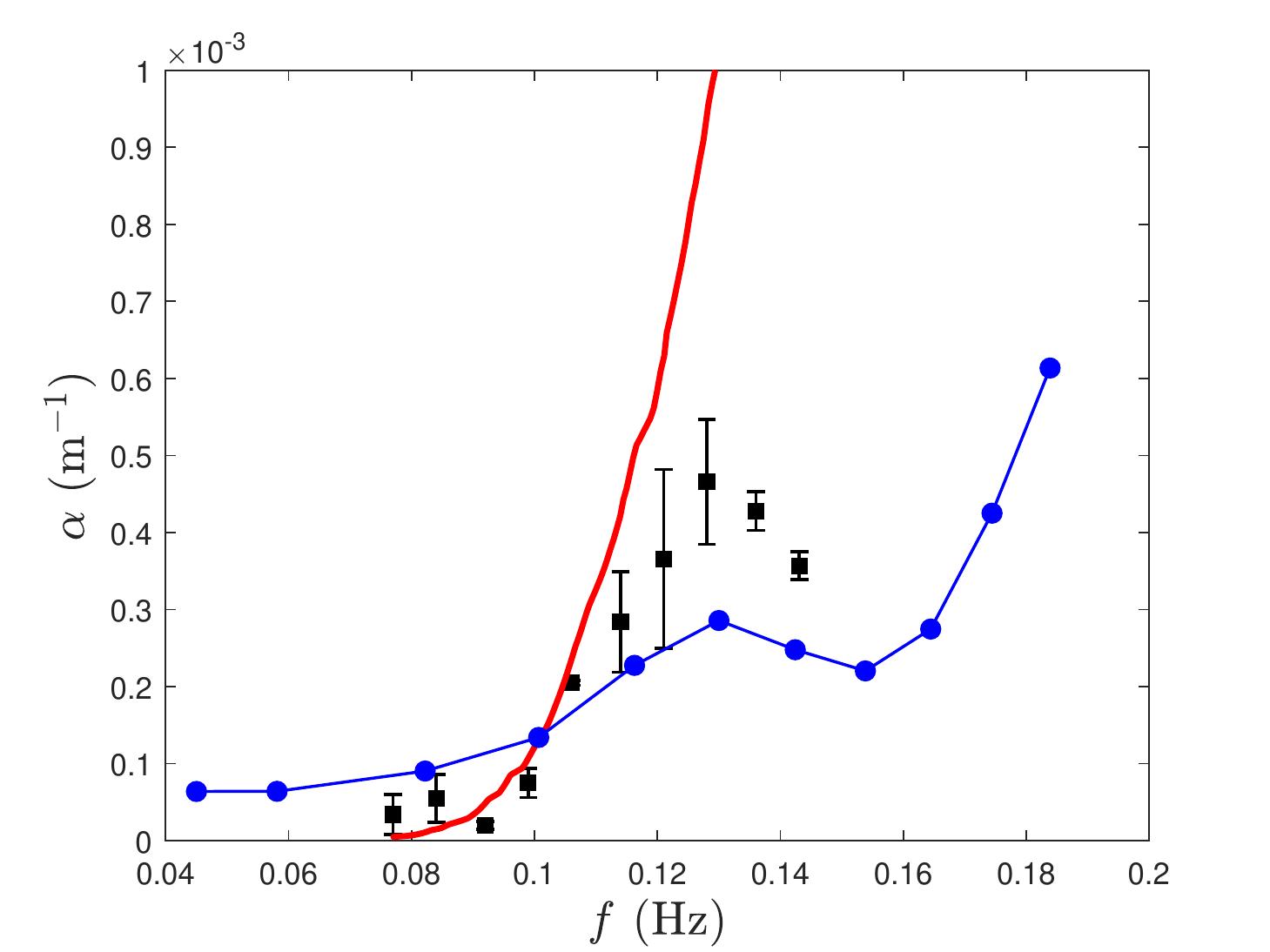}
\caption{Comparison of attenuation rate vs. frequency between measurements (black squares) 
from the 7 February 1983 Bering Sea experiment \cite{wsgcm88} and our closest estimates (blue dots)
for $(\varepsilon,N_i,L_i) = (0.2,124,61.2 \mbox{ m})$.
Predictions (red line) from the linear scattering model of Kohout and Meylan \cite{km08} are also presented.}
\label{bering}
\end{figure}

Figure \ref{bering} shows a comparison with observations from the 7 February 1983 Bering Sea experiment where the roll-over 
is even more pronounced than in the previous situation due to the larger number of data points and smaller experimental errors.
This phenomenon also peaks around $f = 0.13$ Hz here but its amplitude is about twice as high as that recorded in the Greenland Sea.
No experimental data are available past the downslope for $f > 0.15$ Hz.
Among our simulations, the closest results to these measurements follow again from $(\varepsilon,N_i,L_i) = (0.2,124,61.2 \mbox{ m})$
which yields the tallest roll-over as shown in Fig. \ref{compare_all}.
Confronting Fig. \ref{bering} to Fig. \ref{greenland}, similar comments can be made but the agreement is now less satisfactory at the roll-over peak.
While its location remains well echoed, its amplitude is clearly underestimated by the numerical predictions.
The largest value of $\alpha$ from our computations (attained at $f = 0.18$ Hz and attributed to strong scattering) 
still exceeds the roll-over amplitude from these field observations.

The notable discrepancy between the computed and measured peaks at $f = 0.13$ Hz in this case
may be explained by the difference in floe configuration.
The 1983 Bering Sea experiment was conducted as part of the MIZEX West study. 
In their model comparison to this data set, Kohout and Meylan \cite{km08} estimated the ice concentration to be ${\cal C} = 0.72$,
and Perrie and Hu \cite{ph96} assumed the typical floe size to be ${\cal L}_i = 14.5$ m
(Wadhams et al. \cite{wsgcm88} reported a floe diameter of about ${\cal L}_i = 20$ m).
Based on this information, the ice cover during that expedition seemed to consist of relatively small floes
densely packed together, while the ice field specified in our best-matching simulations is sparser with larger floes 
($C = 0.31$, $L_i = 61.2$ m).
It would be of interest to perform simulations with more closely distributed floes of smaller size ($C > 0.31$, $L_i < 61.2$ m)
and see if this might better compare to the Bering Sea measurements,
but we anticipate that such conditions would be more difficult to implement and would entail a higher computational cost
because they may require using a finer grid.
The high ice concentration ${\cal C} = 0.72$ from these field observations also suggest that dissipative mechanisms
not represented in our numerical model might account for the discrepancy with scattering results.

Despite uncertainty in these findings, partly due to the limited number of cases that we investigated, some comments are in order.
The fact that our best match to both experiments is the same set of numerical estimates for $\alpha$
corresponding to the parameter regime $(\varepsilon,N_i,L_i) = (0.2,124,61.2 \mbox{ m})$
would be consistent with the evidence that both data sets show a roll-over at about the same frequency.
However, we must admit that the ice conditions were certainly different during these two separate experiments
and so this observation about the roll-over occurring at the same frequency suggests that
the underlying mechanism may be independent of the floe configuration.
In particular, our assessment demonstrates that the roll-over effect is likely not a resonance peak characteristic of wave scattering by ice floes.
Another related question is as to whether wind input and nonlinear energy transfer would be 
important factors in the roll-over development as suggested by Wadhams et al. \cite{wsgcm88} and other numerical studies \cite{lkdwgs17,ph96}.
Our computations clearly indicate that external forcing is not required to produce a non-monotonic behavior of decay rate
as a function of frequency.
We directly simulated Eqs. \eqref{eq-eta}--\eqref{eq-xi} given initial conditions and did not consider any contributions 
from such driving mechanisms as the wind or a moving load \cite{gp12,gp14,mvbw11}.

The possible role of nonlinearity in this phenomenon is a relevant question considering the relatively high steepness
($\varepsilon = 0.2$) of the incident waves in our best-matching simulations.
A related comment was made earlier when discussing all the various cases displayed in Fig. \ref{compare_all}.
To address this point, we also include in Figs. \ref{greenland} and \ref{bering} 
predictions from the linear scattering model of Kohout and Meylan \cite{km08} which shares some similarities with the present formulation.
These results are extracted from Figs. 8 and 11 of their paper where they were compared to the same Bering and Greenland Seas data.
Appropriate rescaling is applied to convert their dimensionless energy rates per floe number in those figures
to dimensional rates of spatial attenuation.
Kohout and Meylan \cite{km08} tackled the two-dimensional linear problem by looking for plane wave solutions 
in terms of an eigenfunction expansion.
Individual floes of the ice cover were discretely resolved and described as thin elastic plates with distinct properties.
Contributions from the velocity potential were matched exactly at each floe's boundaries
where free-edge conditions were imposed.
Both the length and thickness of each floe were specified randomly about some mean values.

Despite being averaged over $100$ realizations, the corresponding estimates appear to be a bit more fluctuating than 
the other curves in Figs. \ref{greenland} and \ref{bering}, but we can discern a general trend that is monotonically increasing with frequency.
This general trend is approximately linear in Fig. \ref{greenland} while it looks more convex in Fig. \ref{bering}.
It agrees well with both sets of experimental data at low frequencies.
However, as stated in \cite{km08}, Kohout and Meylan's scattering model is unable to reproduce the roll-over in these situations.
Similar to our computations, their linear predictions compare less favorably with the Bering Sea data,
which these authors attributed to their model being less suitable for higher ice concentrations and shorter floe lengths.

\begin{figure}[t!]
\includegraphics[width=\linewidth]{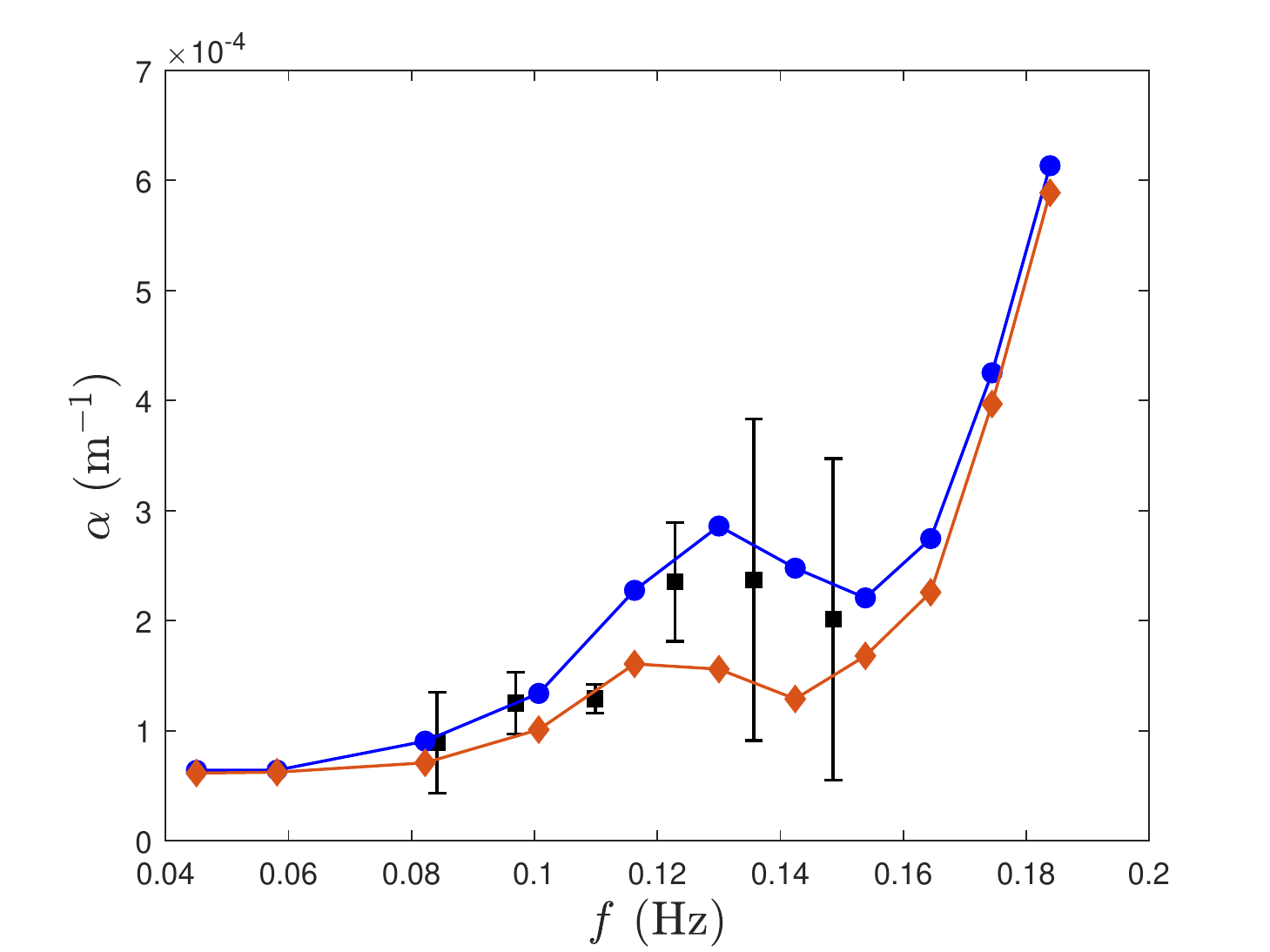}
\caption{Comparison of attenuation rate vs. frequency between measurements (black squares) 
from the 10 September 1979 Greenland Sea experiment \cite{wsgcm88} and our closest nonlinear estimates (blue dots)
for $(\varepsilon,N_i,L_i) = (0.2,124,61.2 \mbox{ m})$.
Predictions (orange diamonds) from the linear version of our numerical model for the same parameter values are also presented.}
\label{greenland_linear}
\end{figure}

As an additional test about the role of nonlinearity, we repeated our best-matching simulations with
$(\varepsilon,N_i,L_i) = (0.2,124,61.2 \mbox{ m})$ but switched all the nonlinear terms off in \eqref{eq-eta}--\eqref{eq-xi}
to explore the propagation and attenuation of wave groups in sea ice from the linear time-evolving viewpoint.
For this purpose, the linear versions of $P_1$ and $P_2$ in \eqref{pressure} are prescribed with
\[
G(\eta) \xi = G_0 \xi \,, \qquad 
\partial_s^2 \kappa + \frac{1}{2} \kappa^3 = \partial_x^4 \eta \,.
\]
The same incident wave conditions and numerical parameters as previously are used in this test.
Figure \ref{greenland_linear} plots our linear results on $\alpha$ together with the Greenland Sea data
which were found to be well emulated by our nonlinear estimates earlier (Fig. \ref{greenland}).
We can still detect a hump but it is much less prominent than its nonlinear or measured counterpart,
and its location seems to be slightly downshifted to near $f = 0.12$ Hz.
Although the respective mechanisms are likely unrelated, this frequency shift between the linear and nonlinear roll-overs 
is reminiscent of a common observation from models for wave propagation over sandbars,
where the peak of Bragg resonance in the reflection coefficient occurs at two slightly different wavenumbers
for the linear and nonlinear predictions \cite{gn07}.
In that context however, the linear peak is upshifted in wavenumber (rather than downshifted) relative to the nonlinear peak.
Aside from the roll-over, we see that the fast increase of $\alpha$ at high frequencies ($f > 0.15$ Hz) 
remains a striking feature of Fig. \ref{greenland_linear},
which confirms the dominant contribution from scattering to wave attenuation in broken floe fields.
While we have not elucidated the exact cause for the roll-over phenomenon, 
there is supporting evidence from our analysis that nonlinear interactions play a major role
in its emergence as revealed by field observations of Wadhams et al. \cite{wsgcm88}.

\begin{figure}[t!]
\includegraphics[width=\linewidth]{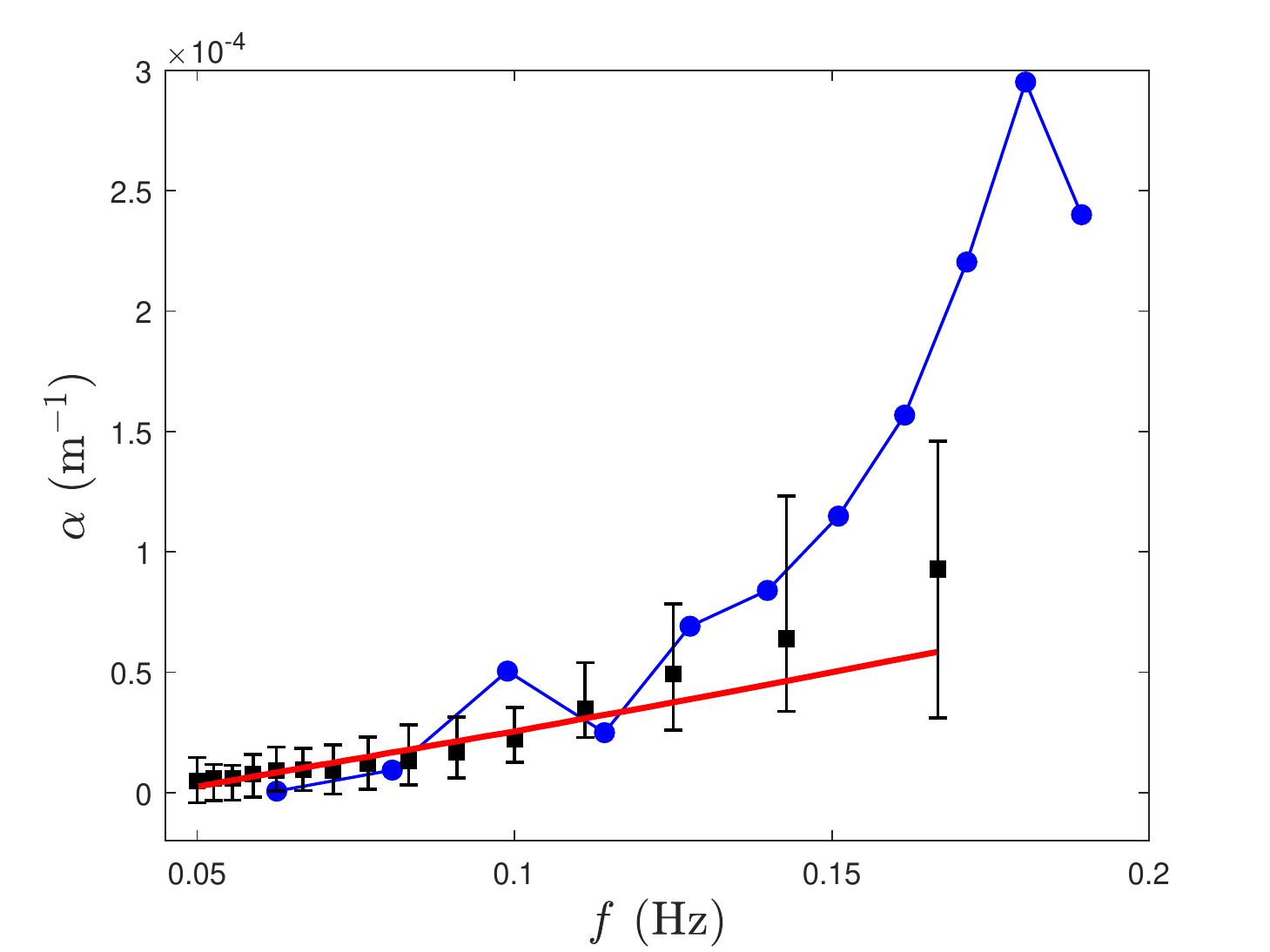}
\caption{Comparison of attenuation rate vs. frequency between measurements (black squares) 
from the Antarctic MIZ \cite{kwdm14,mbk14} and our closest estimates (blue dots)
for $(\varepsilon,N_i,L_i) = (0.05,20,917.3 \mbox{ m})$.
Predictions (red line) from the linear viscoelastic model of Mosig et al. \cite{mms15} are also presented.}
\label{antarctic}
\end{figure}

We now examine a more recent data set that was garnered from the Antarctic MIZ as part of the Australian Antarctic Division's
second Sea Ice Physics and Ecosystem Experiment in 2012.
Wave measurements were acquired simultaneously using advanced sensors at up to five locations along a transect stretching up to $250$ km.
They were included in a preliminary report by Kohout et al. \cite{kwdm14} to support the claim that wave activity and ice extent are correlated.
Meylan et al. \cite{mbk14} conducted a spectral analysis on these data to glean the dependence of decay rates on wave periods.
Figure \ref{antarctic} compares the resulting data on $\alpha$ (extracted from Fig. 8 in \cite{mms15})
with the closest estimates from our nonlinear simulations,
which turns out to be for $(\varepsilon,N_i,L_i) = (0.05,20,917.3 \mbox{ m})$ in this situation.
The latter is different from our best match in the two previous environments, partly because the values of $\alpha$ 
are overall lower here and the roll-over is absent (at least from the experimental data).
The simulated range of frequencies slightly exceeds again the measured range, especially at high frequencies
(however the measured range is better resolved at lower frequencies).
This data set exhibits a monotonically increasing trend for $\alpha$ as a function of $f$,
which is reasonably well reproduced by our numerical predictions (being within or near experimental error).
The computed $\alpha$ tends to rise faster than the measured one as $f$ increases,
and continues in this fashion past $f = 0.17$ Hz where experimental data are not available.

Figure \ref{antarctic} also suggests the presence of peaks for the computed $\alpha$ around two separate frequencies $f = 0.10$ Hz and $0.18$ Hz.
However, the respective wavelengths $\lambda_0 = 160$ m and $48$ m are significantly shorter than the floe size $L_i = 917$ m 
specified in our best-matching simulations, so it is unlikely that these two peaks are indicative of strong scattering.
Furthermore, unlike the previous cases, they would not represent a roll-over effect 
because each of these peaks seems to only involve a single data point.
Looking back at Fig. \ref{compare_all} and noting the substantial difference in $\alpha$ magnitude 
between e.g. $(\varepsilon,N_i,L_i) = (0.05,124,61.2 \mbox{ m})$ and $(0.05,20,917.3 \mbox{ m})$, 
we deduce that small variations in the computed $\alpha$ around these two frequencies
as depicted in Fig. \ref{antarctic} might just be fluctuations possibly due to numerical errors.

Four estimates of ice concentration ${\cal C} = 0.210$, $0.481$, $0.498$, $0.576$ in areas of the Antarctic MIZ
where the sensors drifted are reported in \cite{mbk14}.
These values were estimated using Nimbus-7 scanning multichannel microwave radiometer 
and Defense Meteorological Satellite Program (DMSP) Special Sensor Microwave/Imager Sounder (SSMIS) Passive Microwave Data.
Photographs taken by a camera installed on the upper deck of the ship can be found in \cite{mbk14}
and reveal a rather dense ice cover during this expedition,
with dominant floe sizes ${\cal L}_i$ ranging from a few meters to greater than $100$ m.
While the simulated ice concentration $C = 0.75$ (corresponding to $N_i = 20$ and $L_i = 917.3 \mbox{ m}$)
is higher than the average ${\cal C} = 0.441$ of the four experimental estimates,
we may deem these ice conditions to be similar qualitatively.

The particularly low level of nonlinearity ($\varepsilon = 0.05$) associated with our best match here
may explain why a better agreement is found at lower rather than higher frequencies as mentioned above.
Indeed, Meylan et al.'s spectral analysis of these Antarctic MIZ data suggests that the linear approximation 
would be suitable at low frequencies.
At high frequencies, their investigation identifies a transition where nonlinear processes are believed to become significant.
What exactly provokes this transition in the Antarctic MIZ is unclear,
nonetheless this fact is consistent with the tendency for our weakly nonlinear results on $\alpha$ 
to deviate from the measurements and rise faster as $f$ increases.

As an independent reference, Fig. \ref{antarctic} also includes predictions from the so-called EFS model proposed by Mosig et al. \cite{mms15}.
Like the measurements, these theoretical estimates are extracted from Fig. 8 in \cite{mms15} 
and converted from decay rates for the wave amplitude to ones for the wave spectrum (via multiplication by a factor of $2$).
The EFS model is based on linear thin-plate theory but, unlike the scattering model of Kohout and Meylan \cite{km08},
it was introduced as a continuum viscoelastic formulation in terms of effective parameters.
Because such a representation is of different nature from the floe-resolving approach advocated in the present study,
we do not further describe it here and refer the reader to \cite{mms15} for more details.
Results in \cite{mms15,xg22} show that the EFS model provides a good approximation to these Antarctic MIZ data,
especially at low frequencies where it outperforms other viscoelastic formulations.
For the purposes of our comparison, it should be kept in mind that the EFS curve was obtained as the best fit 
according to a rigorous minimization procedure \cite{mms15}, while this is not so for our numerical estimates.

We see in Fig. \ref{antarctic} that the linear and nonlinear predictions are relatively close together
over the low-frequency region $f < 0.13$ Hz where they both agree well with the field data.
However, they tend to deviate from each other as $f$ increases, with the EFS model slightly underestimating $\alpha$ at high frequencies 
while our nonlinear results overestimate $\alpha$ and exhibit a sharp increase, as discussed above.
We point out that the close fit by the EFS solution was achieved at the expense of excessively large values of shear modulus 
$\mu = 10^{12}$ Pa and kinematic viscosity $\zeta = 10^7$ m$^2$ s$^{-1}$ 
for the effective medium representing the Antarctic MIZ \cite{mms15}.
By contrast, our simulations ignore dissipative processes and assume a more common value of shear modulus for sea ice, i.e.
\[
\mu = \frac{E}{2 (1 + \nu)} = 2.3 \times 10^9 \mbox{ Pa} \,,
\]
based on the characteristic values of Young's modulus and Poisson's ratio given in Sec. 2.1.

\section{Conclusions}

The process of wave attenuation in the MIZ remains a challenging problem due to the heterogeneous nature of this environment
and the presence of various types of sea ice.
Scattering is believed to be the main contributor to wave attenuation in broken floe fields when the wavelength is on the order of the floe size.
We perform direct phase-resolved simulations to investigate this mechanism from a deterministic, evolutionary and nonlinear viewpoint,
in the absence of dissipative effects.
As an extension of the work of Guyenne and P\u ar\u au \cite{gp17}, the numerical model solves the full time-dependent equations
for nonlinear potential flow, combined with a nonlinear bending force to describe the ice cover 
according to the special Cosserat theory of hyperelastic shells.
A lower-dimensional reduction of the governing equations is accomplished by introducing the DNO,
and a high-order spectral method is adopted to solve this reduced system in terms of boundary variables.
Efficient and accurate simulations are achieved by using a series expansion of the DNO together with the fast Fourier transform.
A mixed continuum-piecewise adjustment of the original plate formulation enables the numerical algorithm
to mimic a spatial distribution of ice floes, for which irregular samples can be generated via a randomization procedure.

Focusing on the two-dimensional deep-water case, we examine the scattering and attenuation of incident wave groups
over long distances of propagation, as this is of particular relevance to wave-ice interactions in the ocean.
Various wave regimes and floe configurations are considered, spanning a range of wave steepnesses, wave frequencies
and ice concentrations (with separate levels of ice fragmentation).
For this purpose, computations are run for a very large domain over an extended period, using fine resolutions in both space and time.
Due to this variety of wave and ice conditions, care is taken to include inertial properties of the ice cover and simulate their nonlinear transient effects.

For each set of parameter values, we estimate the spatial attenuation rate as a function of wave frequency,
which is obtained by least-squares fitting an exponential function to a version of the local wave spectrum
(i.e. ensemble-averaged time-averaged squared elevation for each grid point in the ice field).
To do so, results from multiple random realizations of each floe configuration are collected and averaged.
Overall, we find that the decay rate tends to increase with frequency, which is consistent with typical predictions from linear theory.
The sparser the floe field or the lower the incident wave steepness, the faster this increase,
which suggests that the combined effects of scattering due to wave resonance with floes of comparable size 
and successive wave reflections in the presence of multiple floes induce strong wave attenuation.

Interestingly, our attenuation curves also exhibit a non-monotonic behavior which peaks at some frequency
that is not characteristic of strong scattering.
This phenomenon is most apparent in our results for highly fragmented ice covers.
By comparing with field data, we show that it bears a resemblance to the roll-over effect as reported in Arctic MIZ measurements.
Our closest predictions to these data sets correspond to the simulated case with a highly broken floe field and steep incident waves.
This resemblance is particularly striking when compared to the Greenland Sea data.
We confirm that nonlinear interactions play an important role in the roll-over development,
as suggested by previous numerical studies.
However, we find no evidence that external forcing such as wind input is needed.

We are also able to obtain qualitatively good agreement in comparison with Antarctic MIZ measurements,
for which there is no sign of roll-over.
Because decay rates in this case are generally lower than those observed in the Arctic MIZ,
our closest estimates come out of simulations for a dense floe field and gentle incident waves.
Aside from the overall similitude with these sets of field measurements in their range of frequencies, 
our numerical results provide some extrapolation at higher frequencies where data are not available.
They behave favorably with respect to predictions from existing linear theories in these specific cases.
They support the general consensus that scattering is the dominant mechanism for wave attenuation 
when sufficiently large floes are prevalent in the MIZ.

In the future, it would be of interest to extend this numerical work by including dissipation in the nonlinear model to examine whether 
we can emulate the much stronger wave attenuation as typically observed in grease or pancake ice fields \cite{desanti18,nm99,rogers16}.
We may use the same type of parameterization for wave dissipation based on effective viscosity as proposed in \cite{gp17b,lhv91,w87}.
The contribution from dissipative processes to wave attenuation in sea ice and to the roll-over phenomenon in particular remains unclear,
and thus requires further consideration.
Another possible extension would be to investigate the three-dimensional problem with fragmented sea ice and compare with field data. 
Discrepancies that we have observed may partly be attributed to such effects.
The present approach for simulating wave-ice interactions is readily applicable to three dimensions \cite{xg09}.

\clearpage

\clearpage

\clearpage

\clearpage

\clearpage

\clearpage

\clearpage

\clearpage

\clearpage

\clearpage

\clearpage

\begin{thebibliography}{00}

\bibitem{bs12}  
L.G. Bennetts, V.A. Squire, 
On the calculation of an attenuation coefficient for transects of ice-covered ocean,
Proc. R. Soc. A 468 (2012) 136--162.

\bibitem{bmf09}
F. Bonnefoy, M.H. Meylan, P. Ferrant,
Nonlinear higher-order spectral solution for a two-dimensional moving load on ice,
J. Fluid Mech. 621 (2009) 215--242.

\bibitem{cgg19}   
H. Chen, R.P. Gilbert, P. Guyenne, 
Dispersion and attenuation in a porous viscoelastic model for gravity waves on an ice-covered ocean,
Eur. J. Mech. B/Fluids 78 (2019) 88--105.

\bibitem{cheng17}
S. Cheng, W.E. Rogers, J. Thomson, M. Smith, M.J. Doble, P. Wadhams, A.L. Kohout, B. Lund, O.P.G. Persson, C.O. Collins III, S.F. Ackley, F. Montiel, H.H. Shen, 
Calibrating a viscoelastic sea ice model for wave propagation in the Arctic fall marginal ice zone,
J. Geophys. Res. 122 (2017) 8770--8793.

\bibitem{crmb15} 
C.O. Collins III, W.E. Rogers, A. Marchenko, A.V. Babanin, 
In situ measurements of an energetic wave event in the Arctic marginal ice zone,
Geophys. Res. Lett. 42 (2015) 1863--1870. 

\bibitem{cgs21}
W. Craig, P. Guyenne, C. Sulem,
Normal form transformations and Dysthe's equation for the nonlinear modulation of deep-water gravity waves,
Water Waves 3 (2021) 127--152.

\bibitem{cs93}
W. Craig, C. Sulem,
Numerical simulation of gravity waves,
J. Comput. Phys. 108 (1993) 73--83.

\bibitem{dd02} 
G. de Carolis, D. Desiderio,
Dispersion and attenuation of gravity waves in ice: a two-layer viscous fluid model with experimental data validation,
Phys. Lett. A 305 (2002) 399--412.

\bibitem{desanti18}
F. de Santi, G. de Carolis, P. Olla, M. Doble, S. Cheng, H.H. Shen, P. Wadhams, J. Thomson,
On the ocean wave attenuation rate in grease-pancake ice, a comparison of viscous layer propagation models with field data,
J. Geophys. Res. 123 (2018) 5933--5948.

\bibitem{dkp19}
E. Dinvay, H. Kalisch, E.I. P\u ar\u au,
Fully dispersive models for moving loads on ice sheets,
J. Fluid Mech. 876 (2019) 122--149.

\bibitem{db13} 
M.J. Doble, J.-R. Bidlot, 
Wave buoy measurements at the Antarctic sea ice edge compared with an enhanced ECMWF WAM: Progress towards global waves-in-ice modeling,
Ocean Model. 70 (2013) 166--173.

\bibitem{gmt21}
J. Gemmrich, T. Mudge, J. Thomson,
Long-term observations of the group structure of surface waves in ice,
Ocean Dyn. 71 (2021) 343--356.

\bibitem{gn07}
P. Guyenne, D.P. Nicholls,
A high-order spectral method for nonlinear water waves over moving bottom topography,
SIAM J. Sci. Comput. 30 (2007) 81--101.

\bibitem{gp12}
P. Guyenne, E.I. P\u ar\u au, 
Computations of fully nonlinear hydroelastic solitary waves on deep water,
J. Fluid Mech. 713 (2012) 307--329.

\bibitem{gp14}
P. Guyenne, E.I. P\u ar\u au, 
Finite-depth effects on solitary waves in a floating ice sheet,
J. Fluids Struct. 49 (2014) 242--262.

\bibitem{gp17}
P. Guyenne, E.I. P\u ar\u au, 
Numerical study of solitary wave attenuation in a fragmented ice sheet,
Phys. Rev. Fluids 2 (2017) 034002.

\bibitem{gp17b}
P. Guyenne, E.I. P\u ar\u au, 
Numerical simulation of solitary-wave scattering and damping in fragmented sea ice,
In Proc. 27th Int. Ocean Polar Engng. Conf. (ISOPE, 25-30 June 2017, San Francisco, USA), pp. 373--380.

\bibitem{hi98}
M. Haragus-Courcelle, A. Ilichev,
Three-dimensional solitary waves in the presence of additional surface effects,
Eur. J. Mech. B/Fluids 17 (1998) 739--768.

\bibitem{h16}
A. Herman,
Discrete-element bonded-particle sea ice model DESIgn, version 1.3a -- model description and implementation,
Geosci. Model. Dev. 9 (2016) 1219--1241.

\bibitem{hcs19}
A. Herman, S. Cheng, H.H. Shen,
Wave energy attenuation in fields of colliding ice floes -- Part 1: Discrete-element modelling of dissipation due to ice-water drag,
The Cryosphere 13 (2019) 2887--2900.

\bibitem{km08}
A.L. Kohout, M.H. Meylan,
An elastic plate model for wave attenuation and ice floe breaking in the marginal ice zone,
J. Geophys. Res. 113 (2008) C09016.

\bibitem{kwdm14}
A.L. Kohout, M.J.M. Williams, S. Dean, M.H. Meylan,
Storm-induced sea ice breakup and the implications for ice extent,
Nature 509 (2014) 604--607.

\bibitem{lkdwgs17}
J. Li, A.L. Kohout, M.J. Doble, P. Wadhams, C. Guan, H.H. Shen, 
Rollover of apparent wave attenuation in ice covered seas,
J. Geophys. Res. 122 (2017) 8557--8566.

\bibitem{lhv91}
A.K. Liu, B. Holt, P.W. Vachon, 
Wave propagation in the marginal ice zone: model predictions and comparisons with buoy and synthetic aperture radar data,
J. Geophys. Res. 96 (1991) 4605--4621.

\bibitem{m03}
J.R. Marko,
Observations and analyses of an intense waves-in-ice event in the Sea of Okhotsk,
J. Geophys. Res. 108 (2003) 3296.

\bibitem{mbk14}
M.H. Meylan, L.G. Bennetts, A.L. Kohout, 
In situ measurements and analysis of ocean waves in the Antarctic marginal ice zone,
Geophys. Res. Lett. 41 (2014) 5046--5051.

\bibitem{mvbw11}
P.A. Milewski, J.-M. Vanden-Broeck, Z. Wang,
Hydroelastic solitary waves in deep water,
J. Fluid Mech. 679 (2011) 628--640.

\bibitem{mw13}
P.A. Milewski, Z. Wang,
Three dimensional flexural-gravity waves,
Stud. Appl. Math. 131 (2013) 135--148.

\bibitem{msb16}
F. Montiel, V.A. Squire, L.G. Bennetts,
Attenuation and directional spreading of ocean wave spectra in the marginal ice zone,
J. Fluid Mech. 790 (2016) 492--522.

\bibitem{mms15}
J.E. Mosig, F. Montiel, V.A. Squire, 
Comparison of viscoelastic-type models for ocean wave attenuation in ice-covered seas,
J. Geophys. Res. 120 (2015) 6072--6090.

\bibitem{nbst20}
F. Nelli, L.G. Bennetts, D.M. Skene, A. Toffoli,
Water wave transmission and energy dissipation by a floating plate in the presence of overwash,
J. Fluid Mech. 889 (2020) A19.

\bibitem{nm99}
K. Newyear, S. Martin,
Comparison of laboratory data with a viscous two-layer model of wave propagation in grease ice,
J. Geophys. Res. 104 (1999) 7837--7840.

\bibitem{odd21}
P. Olla, G. de Carolis, F. de Santi,
Diffusion of gravity waves by random space inhomogeneities in pancake-ice fields. Theory and validation with wave buoys and synthetic aperture radar,
Phys. Fluids 33 (2021) 096601.

\bibitem{osvbck16}
M.D. Orzech, F. Shi, J. Veeramony, S. Bateman, J. Calantoni, J.T. Kirby,
Incorporating floating surface objects into a fully dispersive surface wave model,
Ocean Model. 102 (2016) 14--26.

\bibitem{pd02}
E. P\u{a}r\u{a}u, F. Dias,
Nonlinear effects in the response of a floating ice plate to a moving load,
J. Fluid Mech. 460 (2002) 281--305.

\bibitem{ph96}
W. Perrie, Y. Hu, 
Air-ice-ocean momentum exchange. Part I: Energy transfer between waves and ice floes,
J. Phys. Oceanogr. 26 (1996) 1705--1720.

\bibitem{pt11}
P.I. Plotnikov, J.F. Toland, 
Modelling nonlinear hydroelastic waves,
Phil. Trans. R. Soc. A 369 (2011) 2942--2956.

\bibitem{rogers16}
W.E. Rogers, J. Thomson, H.H. Shen, M.J. Doble, P. Wadhams, S. Cheng,
Dissipation of wind waves by pancake and frazil ice in the autumn Beaufort Sea,
J. Geophys. Res. 121 (2016) 7991--8007.

\bibitem{s19}
H.H. Shen,
Modelling ocean waves in ice-covered seas,
Appl. Ocean Res. 83 (2019) 30--36.

\bibitem{s20}
V.A. Squire,
Ocean wave interactions with sea ice: a reappraisal,
Ann. Rev. Fluid Mech. 52 (2020) 37--60.

\bibitem{tgrca19}
J. Thomson, J. Gemmrich, W.E. Rogers, C.O. Collins, F. Ardhuin,
Wave groups observed in pancake sea ice,
J. Geophys. Res. 124 (2019) 7400--7411.

\bibitem{thmkk21}
J. Thomson, L. Ho\u sekov\'a, M.H. Meylan, A.L. Kohout, N. Kumar,
Spurious rollover of wave attenuation rates in sea ice caused by noise in field measurements,
J. Geophys. Res. 126 (2021) e2020JC016606.

\bibitem{wsgcm88}
P. Wadhams, V.A. Squire, D.J. Goodman, A.M. Cowan, S.C. Moore,
The attenuation rates of ocean waves in the marginal ice zone,
J. Geophys. Res. 93 (1988) 6799--6818.

\bibitem{ws10} 
R. Wang, H.H. Shen, 
Gravity waves propagating into an ice-covered ocean: A viscoelastic model, 
J. Geophys. Res. 115 (2010) C06024.

\bibitem{w87} 
J.E. Weber, 
Wave attenuation and wave drift in the marginal ice zone, 
J. Phys. Oceanogr. 17 (1987) 2351--2361.

\bibitem{wbsdb13b}
T.D. Williams, L.G. Bennetts, V.A. Squire, D. Dumont, L. Bertino, 
Wave-ice interactions in the marginal ice zone. Part 2: Numerical implementation and sensitivity studies along 1d transects of the ocean surface, 
Ocean Model. 71 (2013) 92--101.

\bibitem{xg09}
L. Xu, P. Guyenne,
Numerical simulation of three-dimensional nonlinear water waves,
J. Comput. Phys. 228 (2009) 8446--8466.

\bibitem{xg22}
B. Xu, P. Guyenne,
Assessment of a porous viscoelastic model for wave attenuation in ice-covered seas,
Appl. Ocean Res. 122 (2022) 103122.

\bibitem{z68}
V.E. Zakharov, 
Stability of periodic waves of finite amplitude on the surface of a deep fluid,
J. Appl. Mech. Tech. Phys. 9 (1968) 190--194.

\bibitem{zs18}
X. Zhao, H.H. Shen, 
Three-layer viscoelastic model with eddy viscosity effect for flexural-gravity wave propagation through ice cover, 
Ocean Model. 131 (2018) 15--23.

\end{thebibliography}
\end{document}